\documentclass[reprint,onecolumn,
amsmath,amssymb,aps]{revtex4}
\usepackage{epsfig,bm,epsf,graphics}
\usepackage{graphicx}% Include figure files
\usepackage{dcolumn}% Align table columns on decimal point
\usepackage{bm}% bold math
\usepackage{xcolor}
\begin{document}
%\draft
%\preprint{APS/123-QED}
\title{Spin Transport in Magnetically Ordered Systems: Ferromagnets, Antiferromagnets and Frustrated Systems}
\author{Danh-Tai Hoang $^{1}$ and Hung T. Diep $^{2}$\footnote{corresponding author}}
\affiliation{%
$^{1}$ Biological Data Science Institute, College of Science, The Australian National University, Canberra, Australia; danhtai.hoang@anu.edu.au\\
$^2$  Laboratoire de Physique Th\'eorique et Mod\'elisation,
CY Cergy Paris Universit\'e, CNRS, UMR 8089\\
2, Avenue Adolphe Chauvin, 95302 Cergy-Pontoise, France;  diep@cyu.fr\\
 }%

\date{\today}% It is always \today, today,

\begin{abstract}
In this review, we outline the important results on the resistivity encountered by an electron in magnetically ordered materials.  The mechanism of the collision between the electron and the lattice spins is shown. Experiments on the spin resistivity in various magnetic materials as well as theoretical background are recalled.  We focus on our works since 15 years using principally Monte Carlo simulations. In these works, we have studied the spin resistivity in various kinds of magnetic systems ranging from  ferromagnets and  antiferromagnets to frustrated spin systems. It is found that the spin resistivity shows a broad peak at the transition temperature in systems with a second-order phase transition, while it undergoes a discontinuous jump at the transition temperature of a first-order transition.  New results on the hexagonal-close-packed (HCP) antiferromagnet are also shown in extended details for the Ising case in  both the frustrated and non-frustrated parameter regions.
\vspace{0.5cm}
%\begin{description}
%\item PACS numbers: 75.25.-j ; 75.30.Ds ; 75.70.-i \\
%\item Keywords: Quantum Spin-Wave Theory; Green's Function Theory; Frustrated Spin Systems ; Non-Collinear Spin Configurations; Dzyaloshinskii-Moriya Interaction; Phase Transition; Monte Carlo Simulation.
%\end{description}
\end{abstract}

                             % Keywords

%Use showkeys class option if keyword
\maketitle

\section{Introduction}
The resistivity encountered by the displacement of an electron driven by an applied electric field in a material is due to its collisions with the material constituents such as atoms. In general these collisions are caused not by direct contacts but by various potentials, magnetic and electric fields from different  sources.  As a matter of fact, the motion of the electron is slowed down except in the superconducting regime. One can mention many simple examples such as the motion of an electron in a magnet, or in a lattice with vibrating atoms (phonons) under an applied electric field.   

The investigation of the resistivity  is one of the most important tasks in condensed matter physics.  Apart from the desire to understand the mechanisms lying behind the resistivity, the numerous applications using the transport properties of electrons  in electronic devices  have motivated an increasing number of studies, experimentally, theoretically and numerically.  The study of the resistivity has started after the discovery of the electron more than a century ago by  the simple free-electron Drude theory taking into account the relaxation time $\tau$ between two successive collisions due to atoms.  The following relation has been established 
\begin{equation}
\sigma=\frac{ne^2\tau}{m}
\end{equation}
where $\sigma$ is the conductivity, $e$ the electron charge, $\tau$ the electron relaxation time, $m$  the electron mass, and $n$  the number of electrons crossing a unit srface per unit of time.

 Over the years, there has been a large number of more realistic theories of resistivity which take into account different interactions.  Nevertheless, this relation is still valid if  the electron mass $m$ is replaced by its effective mass $m^*$ which includes effects of the interactions of the electron with its environment. The relaxation time $\tau$ should also be modified, it is no more a constant but it depends on the collision mechanism.
In a crystal it is known that the effective mass of the electron can be "heavier" or "lighter" than its mass at rest $m_0$ because it contains the effects of various interactions. This strongly modifies the mobility of the electron in crystals. As for  $\tau$,  it has a strong effect on the temperature dependence of the resistivity.   It has been established that $\tau$ depends on a power of the electron energy. This power depends on the collision process such as collisions with charged impurities, neutral impurities, magnetic impurities, phonons, magnons, etc.  Thus, in a crystal
 the total resistivity $\rho_t(T)$  is a sum of the contributions coming from various collision processes. At low temperature ($T$), it is given by

\begin{equation}\label{rhot}
\rho_t(T)=\rho_0+A_1T^2+A_2T^5+A_3\ln{\frac{\mu}{T}}
\end{equation}
where $A_1$, $A_2$ and $A_3$ are constants. The first term is $T$-independent,  the second term,  proportional to $T^2$, stems from the scattering of the conducting electron at low $T$ by lattice spin-waves. Note that the resistivity caused by a Fermi liquid is also proportional to $T^2$ with another coefficient.   The $T^5$ term which is  observed in metals comes from the diffusion of conduction electrons by atomic vibrations.  However,  the resistivity in metals show a linear-$T$ dependence at high $T$.  The last term expresses the contribution from  the quantum Kondo effect, namely the scattering of  conduction electrons by  magnetic impurities at extremely low $T$.

In this review, we focus our attention on the resistivity $\rho$ due to the spin of the conduction electron  in magnetically ordered materials. For short let us call it the "spin resistivity" hereafter.  This spin resistivity has been widely studied both experimentally and theoretically for more than five decades. The rapid development of the field is due mainly to many applications in spintronics.

  We are interested in magnetic materials which show a  phase transition from a magnetically ordered phase, such as ferromagnets and antiferromagnets,  to the paramagnetic state. 
Let us mention in the following some experiments which have been performed in magnetic materials including metals, semiconductors and superconductors. These experiments carried out on various materials  show different shapes of the spin resistivity around the  phase-transition temperature:  SrRuO$_3$ thin films \cite{Xia}, Ru-doped La$_{0.4}$Ca$_{0.6}$MnO$_3$ \cite{Lu2009},  antiferromagnetic $\epsilon$-(Mn$_{1-x}$Fe$_x$)$_{3.25}$Ge \cite{Du2007}, semiconducting Pr$_{0.7}$Ca$_{0.3}$MnO$_3$ thin films \cite{Zhang}, superconducting BaFe$_2$As$_2$ single crystals \cite{Wang-Chen}, LaFeAsO \cite{McGuire2008} and La$_{1-x}$Sr$_x$MnO$_3$ \cite{Santos2009}.  We see in these works that depending on the nature of the compound, $\rho$ can have a pronounced peak \cite{Matsukura} or a change of its slope, or a curvature change  at the transition temperature $T_C$.  Note that in the last case, one has a maximum of the differential resistivity $d\rho/dT$ \cite{Stishov,Shwerer}.

Theoretically,  the $T^2$ magnetic contribution in Eq. (\ref{rhot})  has been obtained  by Kasuya \cite{Kasuya}  taking into account the scattering of the electron spin by the spin waves at low $T$.  However, at higher $T$, specially in the region of the phase transition of the magnetic lattice, there has been no such clear mechanisms explaining different experimental behaviors of the spin resistivity.  de Gennes and Friedel \cite{DeGennes} have
conjectured that the spin resistivity has the origin in  the spin-spin correlation so it should behave as the magnetic susceptibility. As a consequence,  it should diverge at $T_C$.  However, 
Fisher and Langer \cite{Fisher}, and
Kataoka \cite{Kataoka} have made the observation that the range of spin-spin correlation should not be infinite at $T_C$ due to collisions. This changes  the shape of $\rho$ with respect to the magnetic susceptibility near the phase transition.  Let us mention that the resistivity due to magnetic impurities has been calculated by Zarand et al. \cite{Zarand} as a function of the Anderson's localization length.  This parameter expresses in fact a kind the correlation sphere induced around each impurity.  Their result shows that the resistivity peak depends on this parameter,  thus in agreement with the spin-spin correlation idea.

Note that the spin resistivity depends on the spin orientation of  the environment: the electron encounters less resistance in a ferromagnet with spins parallel to its  spin than in a ferromagnet with spins antiparallel to its spin. Imagine a film composed of three ferromagnetic layers where the middle one is a soft  ferromagnet. In the layer-coupling configuration  $\uparrow-\uparrow-\uparrow$, the movement of an up spin perpendicular to the film encounters a resistance $R_{\uparrow}$. One applies now an external magnetic field in the negative direction, small enough to reverse the spins in the (middle) soft layer: one has the three-layer configuration $\uparrow-\downarrow-\uparrow$. The up spin is found to encounter much difficulty to cross the three layers: the resistance is $R_{\downarrow}$ which is much larger than $R_{\uparrow}$.  This is the phenomenon  Giant-Magneto-Resistance (GMR) discovered in Refs. \cite{Fert1,Fert2,Grunberg} which has many applications in spintronics.

The absence of Monte Carlo (MC) simulation in the literature on the spin resistivity has motivated our works since 2007: we have studied the spin current in a number of systems including ferromagnets \cite{Akabli,Akabli2,Akabli3} and antiferromagnets \cite{Akabli4,Magnin,Magnin3,Magnin2012} by MC simulations. The behavior of $\rho$
 as a function of  $T$ has been shown to be in general agreement with experiments and theories mentioned above. In addition to ferromagnets and antiferromagnets, we have also studied the spin resistivity in frustrated spin systems \cite{Magnin2,Hoang2011}. These systems discovered in the early 80's have been intensively studied. Many unusual properties have been found. The reader is referred to Ref. \cite{DiepFSS} for  reviews on various frustrated spin systems.  In  this paper, we will summarize  the most important aspects and results of works on the spin resistivity in ferromagnets, antiferromagnets and in some frustrated magnets.

In section \ref{method}, we present our generic model and the MC method which we employ to study the spin resistivity. Section \ref{review} is devoted to the presentation of our main MC results since 2007. Comparison with some experiments is made in this section. In section \ref{hcplattice}, we show new results in the case of a hexagonal-close-packed (HCP) crystal where we tune a frustration parameter allowing to study both the non-frustrated case and the frustrated case in the phase space.  Conclusing remarks are given in section \ref{conclu}.

\section{Model and Method}\label{method}
\subsection{Model}

We have investigated the spin resistivity in magnetically ordered materials by using a newly-deviced efficient MC simulation method \cite{Magnin,Magnin3}.  The success of the method was demonstrated when we  studied the semiconducting MnTe where the agreement with experiment is  excellent \cite{Magnin2012}.  This case will be reviewed below. In ferromagnets, we found that the spin resistivity has a high peak at the lattice order-disorder transition tremperature $T_C$.  As said earlier, this anomaly comes from the spin-spin correlation \cite{DeGennes,Fisher,Kataoka} but remains finite at $T_C$.  In antiferromagnets, one observes only a broad maximum \cite{Haas}. In addition to ferromagnets and antiferromagnets, we have investigated  the spin resistivity in the following two frustrated systems: the  face-centered cubic (FCC) lattice with  Ising spins \cite{Magnin2} and the $J_1-J_2$ simple cubic (SC) lattice \cite{Hoang2011}.  We found that that the first-order phase transition in these frustrated systems causes a discontinuity of the spin resistivity at $T_C$.

Let us recall here the model and method which have been used in our early works shown in Refs. \cite{Magnin,Magnin2}: simulations have been carried out to calculate the current of itinerant spins moving in the system under the action of an electric field  $\vec \epsilon$ applied in the $x$ direction.  The itinerant spin $\sigma_i$ carried by a conduction electron interacts with its surrounding lattice spins  inside the sphere of radius $D_1$ centered at its position at the time $t$ on its trajectory across the crystal.  The Hamiltonian is supposed to be

\begin{equation}
{\cal H}_l=- \sum_{j} I_{ij} \vec {\sigma}_i\cdot \vec S_j
\end{equation}
where the sum is carried over all lattice spins in the sphere centered at the itinerant Ising spin $\vec \sigma_i$.  $I_{ij}$ denotes the distance-dependent interaction between $\vec \sigma_i$ and $\vec S_j$.  To be general, we also consider the following interaction between an electron spin with  neighboring conduction spins within a sphere of radius $D_2$

\begin{equation}
{\cal H}_i=- \sum_{j} K_{ij} \vec  {\sigma}_i\cdot \vec  {\sigma}_j
\end{equation}
For simplicity, we suppose the distance-dependent interactions are
\begin{eqnarray}
I_{ij}&=&I_0 \exp(-Br_{ij}) \\
K_{ij}&=&K_0 \exp(-Cr_{ij}) \label{K}
\end{eqnarray}
where $I_0$, $B$, $K_0$ and $C$ are constants to be chosen so that the energy of a conduction electron spin
is much smaller than that of a lattice spin. This choice is made from a physical consideration: in choosing so, we avoid the influence
of itinerant spins on the ordering of the lattice spins.  Note that this choice is justified in the almost-free
electron model where $I_0\simeq K_0\simeq 0$, and in semiconductors where they are larger but still weak with
respect to the exchange intergrals of the lattice spins $J_1$ and $J_2$.  A discussion in details  on the choice of these parameters has been given for example in Ref.  \cite{Magnin}.  

Let us assume in the folklowing  a concentration of one
itinerant electron per two lattice cells.  This concentration is thus of the order of electron concentration in normal metals which is $~10^{23}/cm^3$.  With this concentration, it is obvious that the averaged distance between two conduction electrons is much larger than the cutoff distance $D_2$ which is of the order of the lattice constant.  However, due to the attractive nature of the electron-electron interaction, Eq. (\ref{K}), it is neccessary to introduce a chemical potential term to insure that itinerant spins are uniformmly dispersed in the crystal and they do not form  clusters. This chemical potential is written as 
\begin{equation}\label{chempot}
{\cal H}_p=D[n(\vec r)-n_0]
\end{equation}
 where $D$ is a positive constant, $n(\vec r)$ denotes the concentration of conduction spins in the sphere of  cutoff radius $D_2$ centered at the position $\vec r$ of the conduction spin under consideration, and $n_0$ the averaged electron concentration.  

$I_0$ is defined in Eq. (5) which represents the magnitude of the interaction between a conduction electron and a localized lattice spin [Eq. (3)]. $ K_0$ is the magnitude of the interaction between  two conduction electrons [see Eq. (4) and (6)]. $D$ is the magnitude of the chemical potential [Eq. (7)]. When we simulate a real material, if we have experimental data on the resistivity and the exchange interactions, such as in the case MnTe presented in section \ref{MnTe}, we can estimate the values of these coefficients. 

%We perform the simulation by taking into account the relaxation time of the lattice spins \cite{Magnin3,Magnin4,Hohenberg} which is given by
%
%\begin{equation}\label{tau}
%\tau_L=\frac{A}{|1-T/T_C|^{z\nu}}
%\end{equation}
%where $A$ is a constant, $\nu$  the correlation critical exponent, and $z$ the dynamic exponent.  We see that as $T$ tends to $T_C$, $\tau_L$ diverges.  This phenomenon is known as the critical slowing-down.  For the Ising spin model, $\nu=0.638$ (3D Ising universality) and $z=2.02$ \cite{Prudnikov}.  We have previously shown  that $\tau_L$ strongly  affects the shape of  $\rho$ at $T_C$ \cite{Magnin3}.  By choosing $A=1$, we fix $\tau_L=1$ at $T=2T_C$ deep inside the paramagnetic phase far above $T_C$. This value is what we expect for thermal fluctuations in the disordered phase.

\subsection{Simulation Method}\label{simul}

Let us study a film of  size $N_x\times N_y\times N_z$ where $N_z$ is the film thickness which is much smaller than the sizes $N_x$ and $N_y$ in the $x$ and $y$, respectively.  Usually, we use $N_z=4 - 8$ and $N_x=N_y=20 - 60$. Itinerant spins move  under the action of an electric field acting on the electron charge, applied in the direction $x$. The electric field energy is

\begin{equation}
\mathcal{H}_E  =  -e\vec {\epsilon}\cdot\vec {\ell}
\end{equation}
where $e$ is the electron charge, $\vec  \epsilon $ the applied electrical field and $\vec  \ell$ the displacement vector of the electron.\\

 The interaction between the lattice spins is given by the Hamiltonian
\begin{equation}
\mathcal{H}_L  =  -J\sum_{i,j} \vec {S_i} \cdot \vec {S_j} \label{HL}
\end{equation}
where $J$ is the exchange interaction between nearest neighbors (NN) $\vec S_i$ and $\vec  S_j$. For ferromagnets $J>0$, and for antiferromagnets  $J<0$.

We use the standard Monte Carlo (MC) method to thermalize the lattice alone at $T$. Next, we introduce the electrons into the lattice. We suppost that each electron spin interacts with the surrounding lattice spins inside a sphere  centered at its position,  of radius $D_1$. The electron spin also interacts with other electron spins within a sphere of radius $D_2$.  Electrons move under the applied electric field.

In order to obtain the spin current we have to thermalize the state of the spins of conduction electrons in the lattice. This is done  by the following steps: 

i) we take a conduction spin and calculate its actual energy $E_{old}$ using the different interactions mentioned above, 

ii)  we make a trial move  $\vec \ell$ for the electron in a random direction between 0 and $a$ where $a$ is the lattice constant. If the move  $\vec \ell$ takes the itinerant electron outside the sample, then we put it inside on the other sample end by virtue of the periodic boundary condition,

iii) we then calculate the new energy $E_{new}$. If  $E_{new}<E_{old}$, then the trial position is  accepted. Otherwise, it is accepted with the probability $\exp [ -(E_{new}-E_{old})/(k_BT)]$.

iv) we take another conduction electron and repeat the three steps above.  We continue with other electrons until all electrons are considered: this accomplishes one MC step/spin. 
A large number of MC steps/spin is neccessary to arrive at a steady current state. We next  average physical quantities of interest at the temperature $T$ under consideration.  

iv) we take another $T$ and repeat the above four steps. We should cover the temperature region of interest.  

This paper is a review of our publications over the past 15 years. In each publication, we used various sample sizes to detect finite size effects. We increased the sample  size until the results do not depend anymore on the size. The results are considered as valid thermodynamically. The finite size effect and the finite size scaling are  what the simulators do before reaching conclusions. This problem is particularly important when one wants to study the criticality or the order of a phase transion. In the review, we do not repeat these details which depend on the studied system. Rather, we emphasize on  the results. The reader interested on the details of the simulation methods is referred to each original publication.

We calculate the spin resistivity $\rho$ as :
\begin{eqnarray}
\rho & = & \frac{1}{N_{s}}
\end{eqnarray}
where $N_{s}$  is the number of mobile electrons passing thru a  unit surface perpendicular to the direction of the applied electric field $x$,  per MC  time unit.  An application with a real material using  real units is presented in subsect. \ref{MnTe}.

For  a good thermal average, we have to perform  very long MC runs and we proceed as follows: for each configuration of the lattice spins we average the spin resistivity over $N_1$ MC steps, then  we thermalize again the lattice with $N_2$ MC steps to get rid of the correlation between lattice spin configurations,  before averaging again the resistivity for $N_1$ MC steps.  We repeat this  $N_1+N_2$ cycle for a large number of times $N_3$.  The total MC steps of resistivity averaging is  about  $4\times 10^5$ steps per spin in our runs.   We found by comparison that this "multi-step" averaging method strongly suppresses  statistical fluctuations seen in our earlier work \cite{Akabli2}.  

It is obvious that the larger $N_1$ and $N_2$ yields the better statistics.  We know that depending on $T$, the relaxation time $\tau_L$ of the lattice spins varies. We have to compare  $\tau_L$ with the relaxation time $\tau_I$ of conduction electron spins in order  to choose a right value of  $N_1$ in order to calculate the average of the resistivity with one lattice spin configuration at $T$.   We know the two limiting cases. The first case is when $\tau_L \simeq \tau_I$. In this case, we have to take $N_1=1$, namely the lattice spin configuration should change with each move of itinerant spins. The second case is when $\tau_L \gg \tau_I$. In this case  itinerant spins can be scattered many times along its trajectory across the same lattice configuration and for many times across the lattice.

In order to choose a right value of $N_1$, we consider the following temperature dependence of $\tau_L$ in non frustrated spin systems.  The relaxation time is expressed in this case as \cite{Hohenberg,Peczak,Prudnikov}
\begin{equation}\label{transport_tau}
\tau_L=\frac{A}{|1-T/T_C|^{z\nu}}
\end{equation}
where $A$ is a constant, $\nu$  the correlation critical exponent, and $z$ the dynamic exponent which depends on the spin model and space dimension.  For 3D Ising model, $\nu=0.638$ and $z=2.02$.  From this expression, we see that as $T$ tends to $T_C$, $\tau_L$ diverges.  In the critical region around $T_C$ the system encounters thus the so-called "critical slowing down": the spin relaxation is extremely long due to the divergence of the spin-spin correlation.  When we take into account the temperature dependence of $\tau_L$, the shape of the resistivity is  strongly modified near $T_C$ where $\tau_L$ is very long, in contrast to the paramagnetic phase where the relaxation time is very short due to rapid  thermal fluctuations.  We should emphasize that at low $T$, $\rho$ does not depend on $\tau_L$ since in that temperature range where the ordering of the lattice spins is almost perfect: the spin landscape from one microscopic state to another does not change significantly, so  the motion of the itinerant electron spin does not significantly vary (see Ref. \cite{Magnin3}).

Finally, we note that we  have also used the Boltzmann's equation combined with MC data to study the spin resistivity \cite{Akabli3}. However, the shape of the resistivity peak at the transition temperature does not agree well with experiments, unlike that obtained from direct MC simulations as shown below. This proves the efficiency of MC simulations for the calculation of the spin resistivity in magnetically ordered materials. The present review therefore aims at promoting this method.

\section{Review}\label{review}
\subsection{Spin Resistivity in Ferromagnets and Antiferromagnets}

\begin{figure}[th]
\centering
 \includegraphics[width=80mm,angle=0]{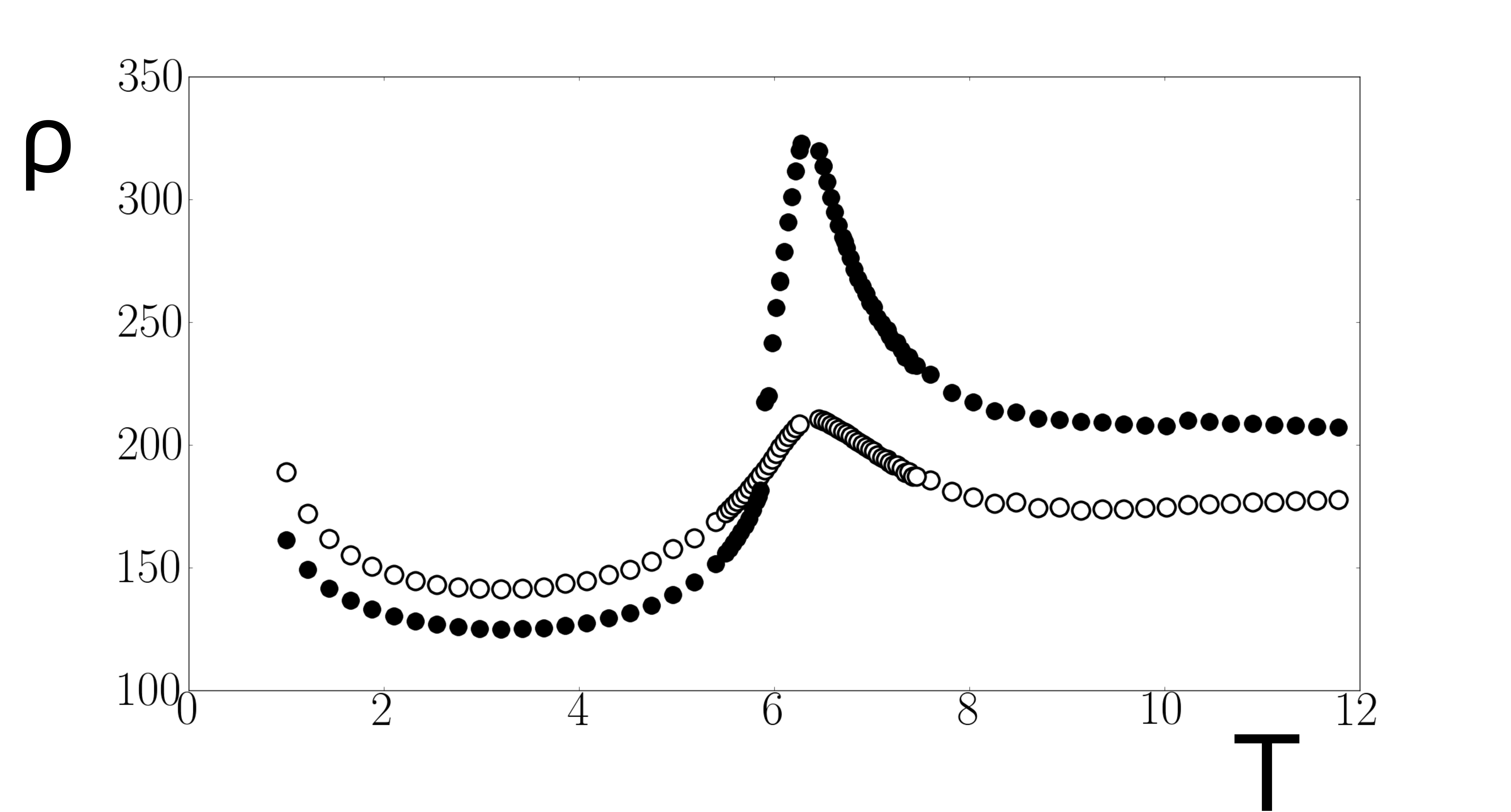}
\caption{ Resistivity $\rho$ as a function of $T$, obtained by simulations using the $T$-dependent relaxation time, Eq. \ref{transport_tau}, for ferromagnet (black circles) and antiferromagnet (white circles) of BCC structure, with arbitrary unit  in zero magnetic field. Other parameters:  $\epsilon=1$, $I_0=2$, $K_0=0.5$, $A=1$.}\label{RX-F-AF}
\end{figure}

Experiments mentioned above amongst numerous other data show that the spin resistivity in ferromagnets has a sharp peak at the transition temperature $T_C$ of the lattice. We know by the theory of phase transitions and critical phenomena that in the region around $T_C$, the so-called "critical slowing-down" phenomenon happens, resulting in extremely large $\tau_L$.  The peak in $\rho$ is due to this phenomenon via Eq. (\ref{transport_tau}) where $\tau_L$ diverges at $T_C$. Our MC simulations using the method described above in the case of a ferromagnet where the lattice spins are of the Ising type show a sharp peak at $T_C$ (see Fig. \ref{RX-F-AF}) in agreement with experimental data. We note  that the spin resistivity for ferromagnets (as well as for antiferromagnets) increases with decreasing $T$ at low $T$. This can be explained by several causes: the freezing or crystallization of conduction electrons takes place at low $T$ so that   just a small number of conduction spins with decreasing $T$ is mobile, or it may be the classical counter effect of the quantum Kondo electron-impurity scattering  if one considers the few excited lattice spins at low $T$ are independent impurities, see the last term of Eq. (\ref{rhot}).   Note that the shape of $\rho$
depends on the lattice type,  interaction strengths encountered by the conduction electrons,  electron concentration, relaxation time, spin model, applied magnetic-field amplitude etc. In Ref. \cite{Magnin}, we have shown that  a decrease in the interaction between conduction spins, $K_0$, reduces the increase of $\rho$ as $T\rightarrow 0$, an applied magnetic field reduces the height of the resistivity peak  and  the larger electron density reduces $\rho$.   Finally, we emphasize that $\rho$ depends on the material  intrinsically via the critical exponents $\nu$ and $z$, see Eq. (\ref{transport_tau}).

If we wish to compare simulated spin resistivity to experimental measurements performed on a given material, we have to use in the simulation the available experimental physical parameters of that material.  An example of quantitative comparison for semiconductor MnTe is shown in  subsection \ref{MnTe}.

%Fig1

Note that the magnetic field applied on the system reduces the peak height as shown in our work Ref. \cite{Magnin},  in agreement with experiments \cite{Lu2009}.

Unlike ferromagnets,  antiferromagnets have not been much studied. 
 Haas \cite{Haas} has shown that  in contrat to ferromagnets where the resistivity $\rho$ has a sharp peak at the order-disorder transition of the lattice spins, in antiferromagnets there is no such a peak.
 Using MC simulations, we found that the peak does exist in an antiferromagnet but it is less sharp compared to that of a ferromagnet, as seen in Fig. \ref{RX-F-AF}.   We think that the alternate change of sign of the spin-spin correlation with distance may have something to do with the absence of a sharp peak.  We have tested this idea on the effect of the cut-off distance $D_1$ \cite{Magnin2}:  in an antiferromagnet, when we increase $D_1$, we will include successively up-spin shells and down-spin shells in the sphere of radius $D_1$.  Consequently, the difference between the numbers of up and down spins in the sphere oscillates with varying $D_1$, giving rise to the oscillatory behavior of $\rho$ observed at small $D_1$, unlike in ferromagnets.  

At this stage, we note that  the presence of an itinerant spin will break the invariance between a ferromagnet and its antiferromagnet counterpart in  the local Mattis transformation ($J_{ij}\rightarrow -J_{ij}, \vec S_{j}\rightarrow -\vec S_{j}$).

\subsection{Frustrated $J_1-J_2$ Model on a Simple Cubic Lattice}
\begin{figure}[th]
\centering
 \includegraphics[width=40mm,angle=0]{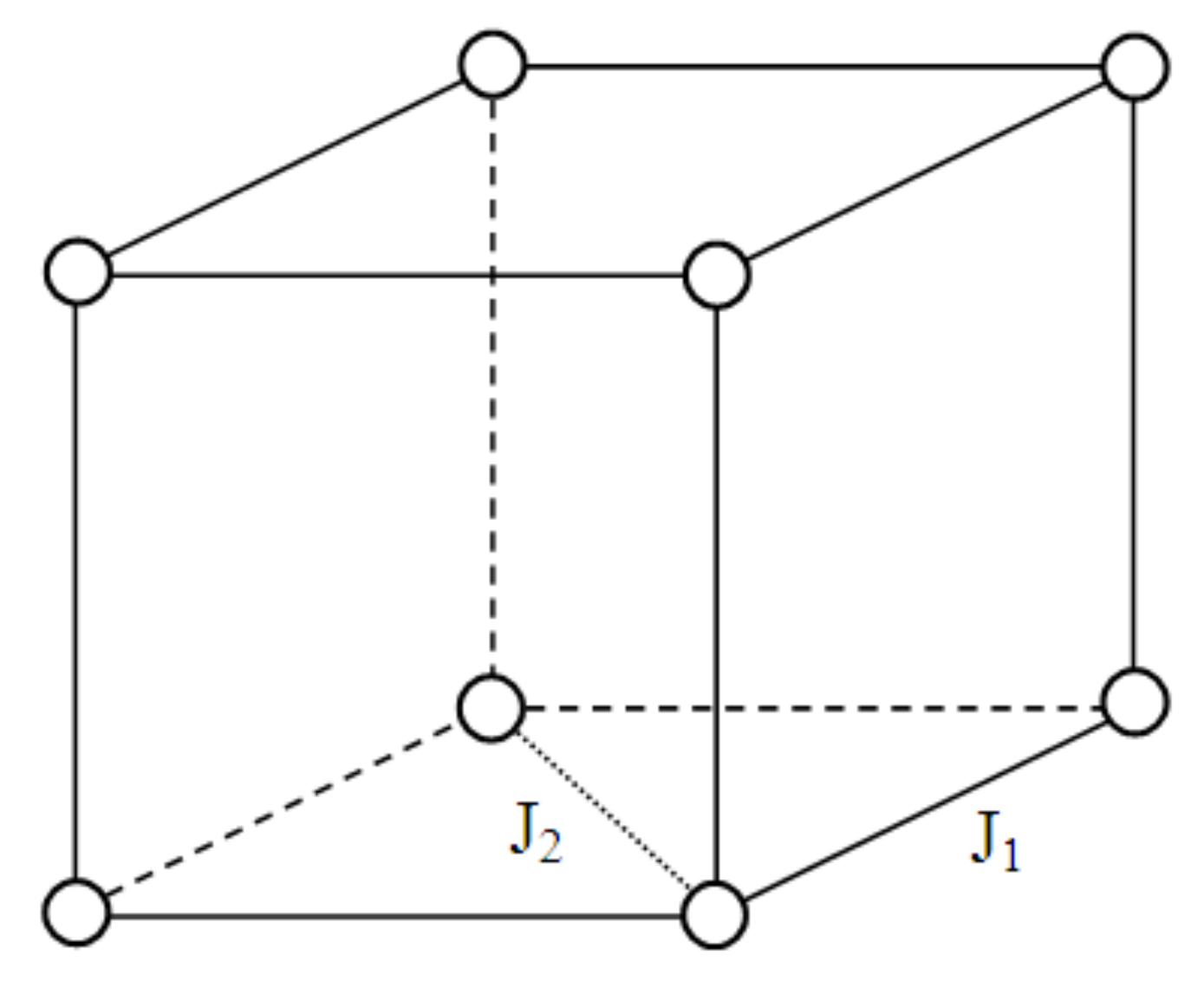}
\caption{Simple cubic lattice where the NN and NNN interactions, $J_1$ and $J_2$, are  indicated.} \label{fig:model}
\end{figure}

Let us  consider the simple cubic lattice with NN and NNN interactions as shown  in Fig. \ref{fig:model}.  The Hamiltonian is written as
\begin{equation}\label{HLJ1J2}
{\cal H} = -J_1\sum_{(i,j)} \vec{S}_i \cdot \vec{S}_j -J_2\sum_{(i,m)} \vec{S}_i \cdot \vec{S}_m
\end{equation}
where the first sum $\sum_{(i,j)}$ is made
over the NN Ising spin pairs  $\vec{S}_i$ and  $\vec{S}_j$ with  interaction $J_1$, and the second sum $\sum_{(i,m)}$ is performed over the NNN pairs with interaction $J_2$.  

\begin{figure}[th]
\centering
 \includegraphics[width=40mm,angle=0]{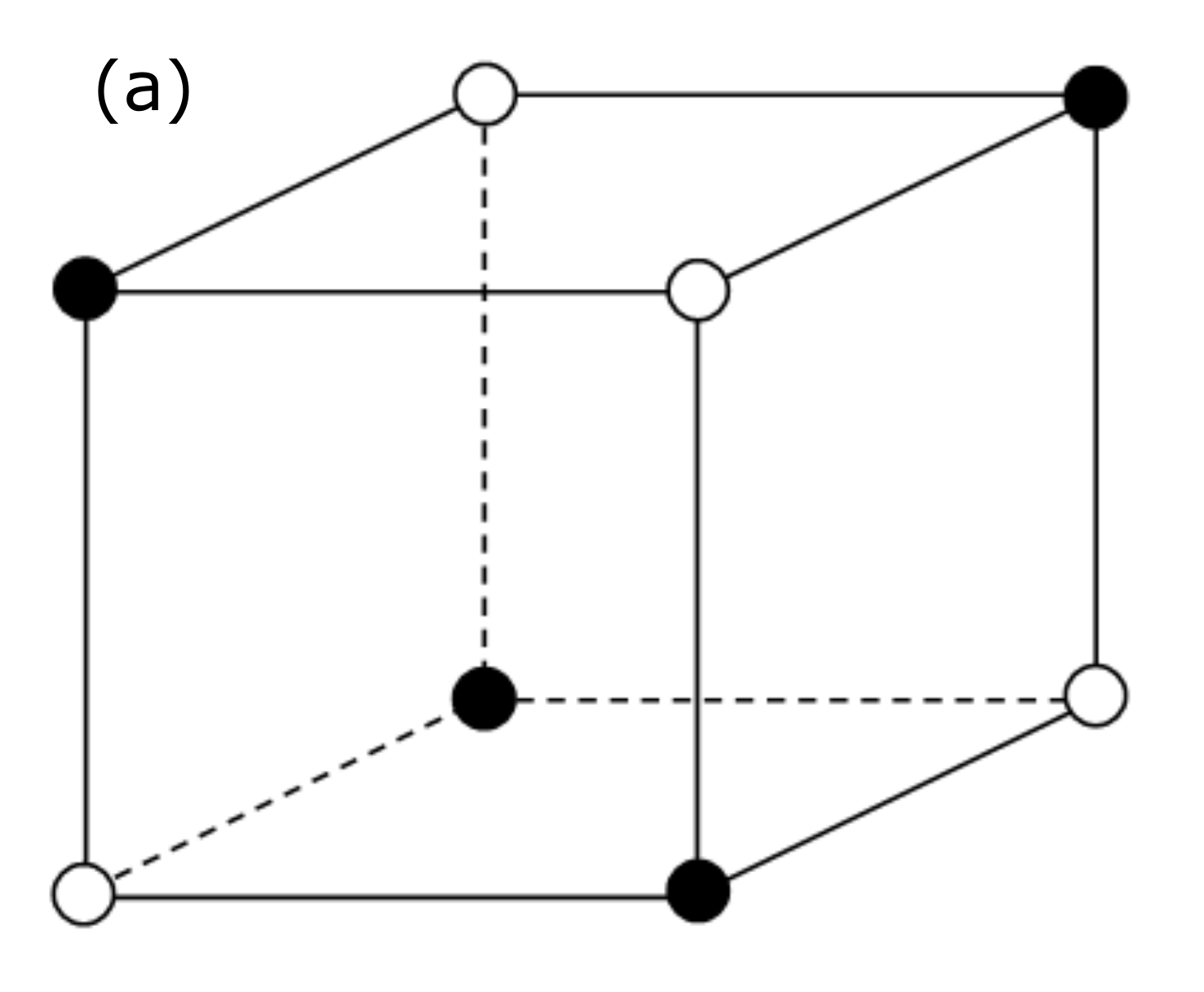}
\includegraphics[width=40mm,angle=0]{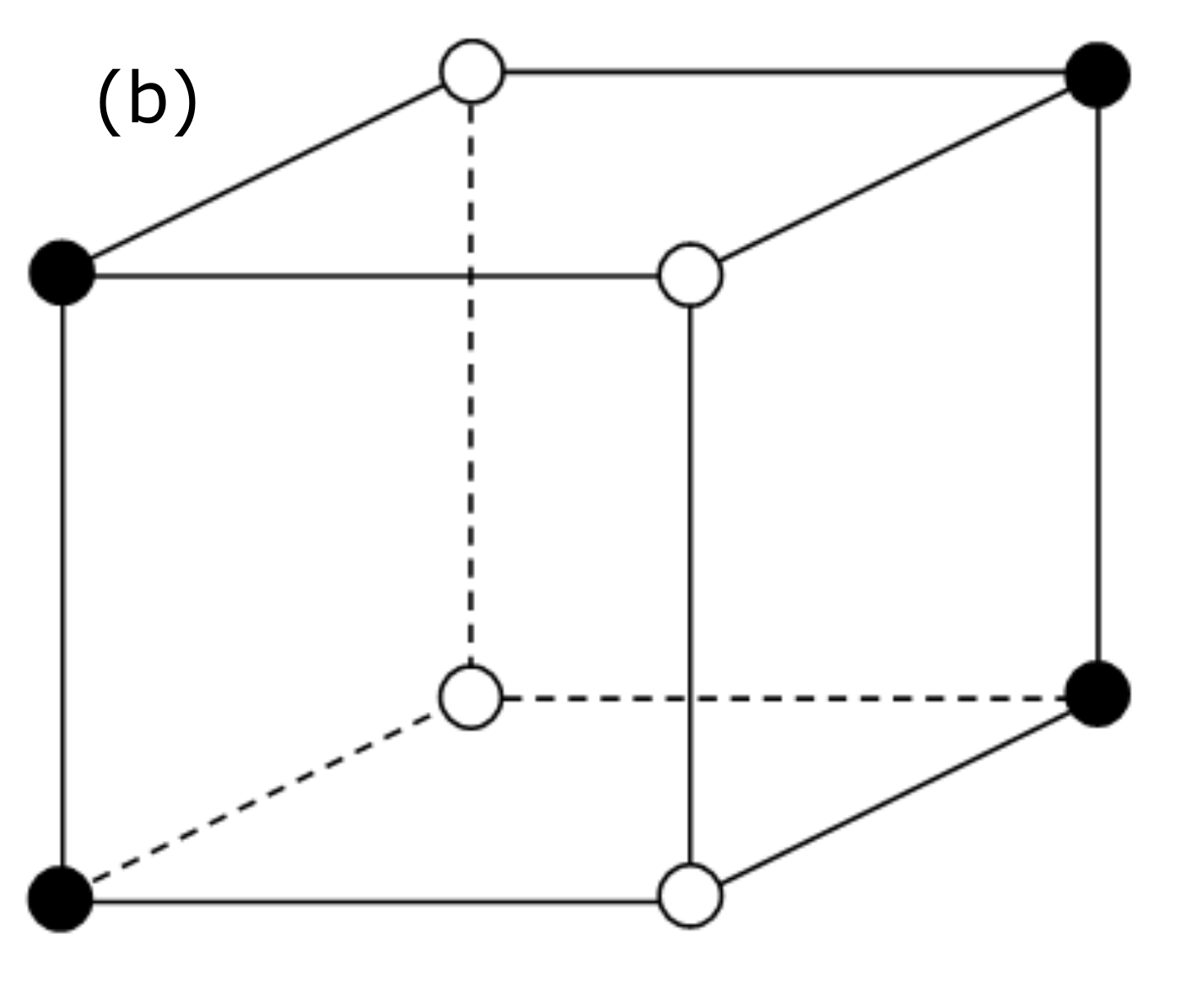}
\caption{Ground state (GS) of the simple cubic lattice with Ising spins: (a) GS when $|J_2|<0.25 |J_1|$; (b) GS when $|J_2|>0.25 |J_1|$. White (black) circles denote up (down) spins. See text for comments. } \label{fig:SC}
\end{figure}

We focus our attention on the region of parameters which gives rise to a frustration.  For that purpose,  we assume that $J_1$ is an antiferromagnetic interaction, namely $J_1=-J<0$ ($J>0$),  and $J_2$ is also antiferromagnetic. We put $J_2=-\eta J$ where $\eta$ is a positive parameter.  The ground state (GS) of this system can be obtained  by minimizing the energy, or by comparing the energies of different spin configurations. We can also numerically minimize the  energy by using the steepest descent method \cite{Ngo2007}.  We obtain the GS antiferromagnetic configuration shown  Fig. \ref{fig:SC}a for $|J_2|<0.25 |J_1|$, and the GS spin configuration shown in Fig. \ref{fig:SC}b for $|J_2|>0.25 |J_1|$.  This latter configuration is 3-fold degenerate because we can choose the parallel NN spins either on $x$, or $y$ or $z$ axis. In addition, with the permutation of black and white spins, we have the total degeneracy equal to 6.

%Fig2

%Fig3

We note in passing that in the case of the Heisenberg model in the frustrated region ($|J_2|>0.25 |J_1|$) the phase transition has been shown to be of first order \cite{Pinettes}.  The system is very unstable due to its large degeneracy. In the case of the Ising spin on the SC lattice treated here, we found that the first-order character of the phase transition is even stronger \cite{Hoang2011}.  

We use $J_1=-J=-1$ (AF interaction) for the coupling between NN lattice spins in the simulations. The energy is thus measured in the unit of $J$ and the temperature is in the unit of $J/k_B$. All distances  ($D_1$ and $D_2$) are in the unit of  the lattice constant $a$.

Simulations have been carried out by using the temperature-dependent relaxation time of the lattice spins given by Eq. (\ref{transport_tau}) where we have taken $A=1$ and $\tau_L=1$ at $T=2T_C$  far from $T_C$. Such a choice leads to $\tau_L=1$ at that temperature expected for fast thermal fluctuations in the paramagnetic phase far above $T_C$.

\begin{figure}[h]
\centering
\includegraphics[width=11cm,angle=0]{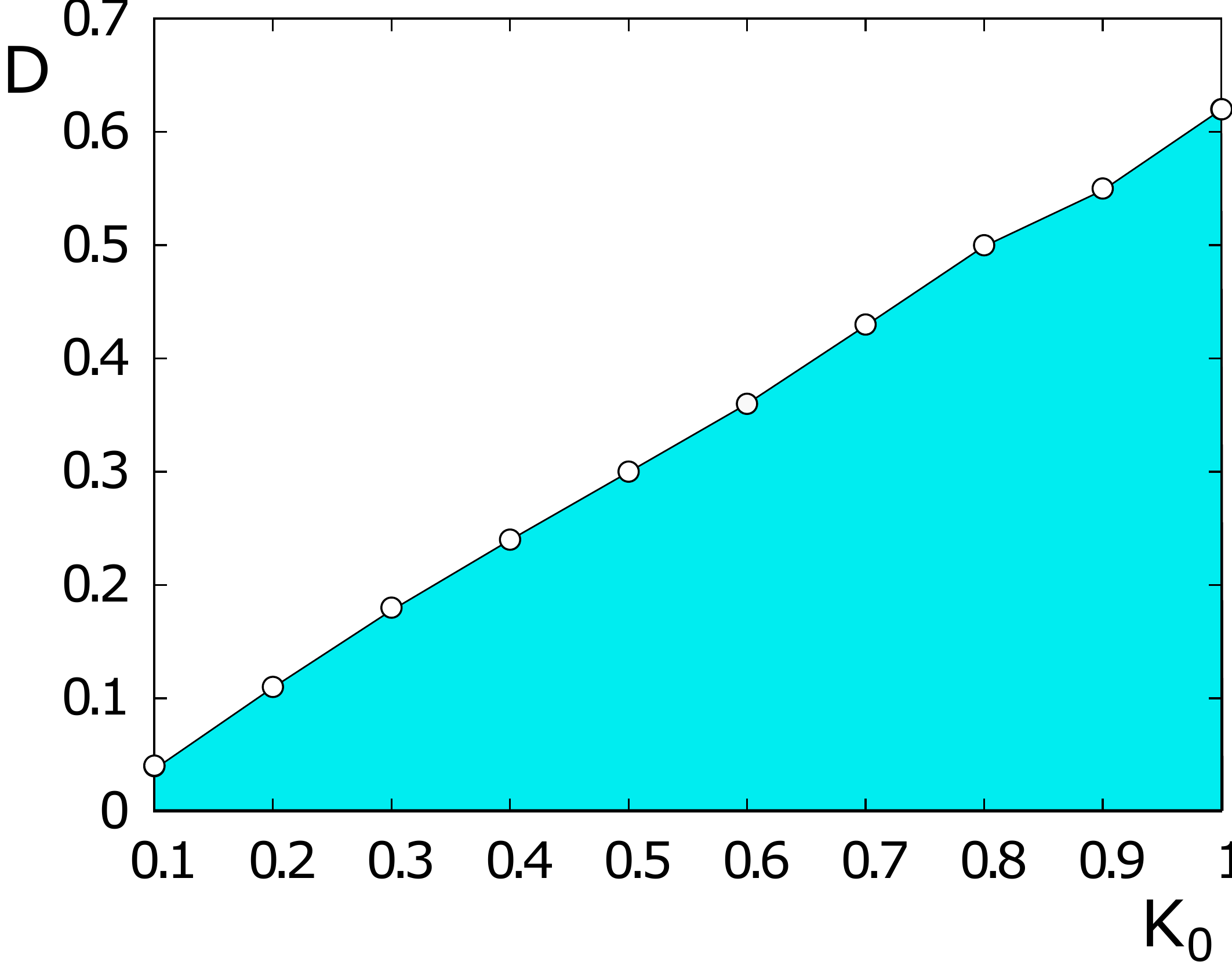}
\caption{$D$ versus $K_0$ in the case where $\eta=0.26$. The collapse region is in green. We have used $D_1=D_2=1$, $I_0=0.5$, $\epsilon=1$.}
\label{fig:J1J2_KD}
\end{figure}

Since we suppose that the interaction between conduction electron spins is attractive,  a chemical potential is required to avoid the collapse of the system, namely to avoid that all conduction spins form a cluster  [cf. Eq.(\ref{chempot})]. The chemical potential in thermodynamics makes the particle uniformly distributed in the space. Its strength is expressed by  $D$ which has to be chosen in accordance with $K_0$.  Figure \ref{fig:J1J2_KD} displays the phase diagram in the space $(K_0,D)$ which shows the collapse region. This allows us to avoid this region and choose an appropriate value of $D$ for a given $K_0$.

%Fig4

To see the effect  of the nature of phase transition on the spin resistivity,  in the following we focus on two typical cases: $\eta=0.2$ and $\eta=0.26$ which belong respectively to the regions of second- and  first-order transition.

\noindent \textbf{A. For $\eta=0.2$:}\\
The spin resistivity at temperatures below $T_N$ oscillates with varying $D_1$.  By analyzing the ratio of the number of up spins to the number of down spins in the sphere of radius $D_1$, we found  that it oscillates with varying $D_1$: the maxima (minima) of $\rho$ correspond to the case of largest (smallest) numbers of parallel (antiparallel) spins in the sphere \cite{Hoang2011,Magnin2}. At very high temperatures where the lattice spins are disordered,  the  numbers of up spins and down spins in the sphere of radius $D_1$ should be  equal. There is however a very small oscillation  if the temperature is close to $T_N$ and if $D_1$ is small.

\begin{figure}[ht]
\centering
\includegraphics[width=10cm,angle=0]{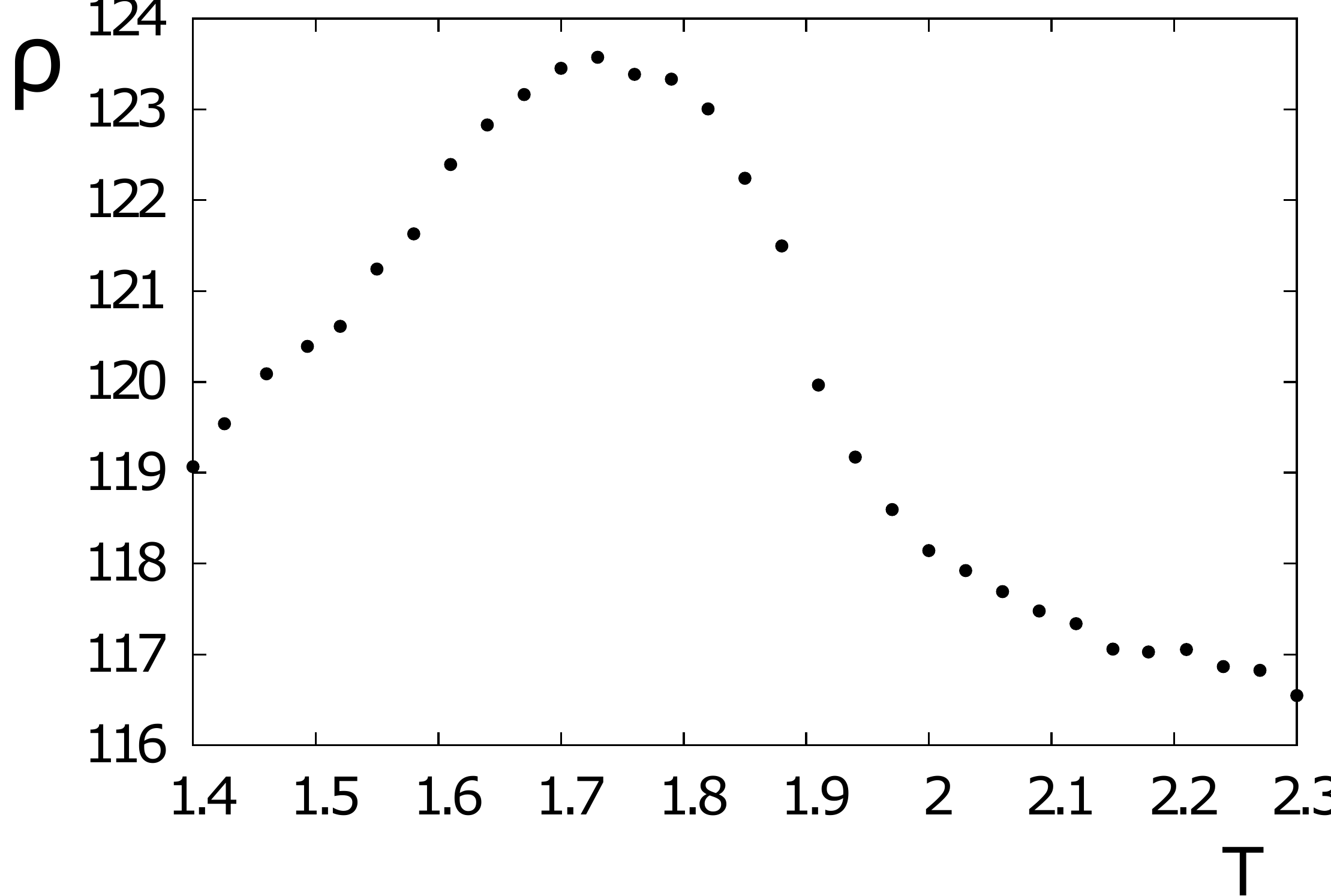}
\caption{Spin resistivity versus $T$ in the case  $\eta=0.2$, $D_1=1.2$.  We have used the lattice size $N_x=N_y=20, N_z=6$. Other variables are $I_0=K_0=0.5$, $D_2=1$, $D=1$, $\epsilon=1$.}
\label{J1J2_R_0.2_1.2}
\end{figure}

\begin{figure}[h]
\centering
\includegraphics[width=15cm,angle=0]{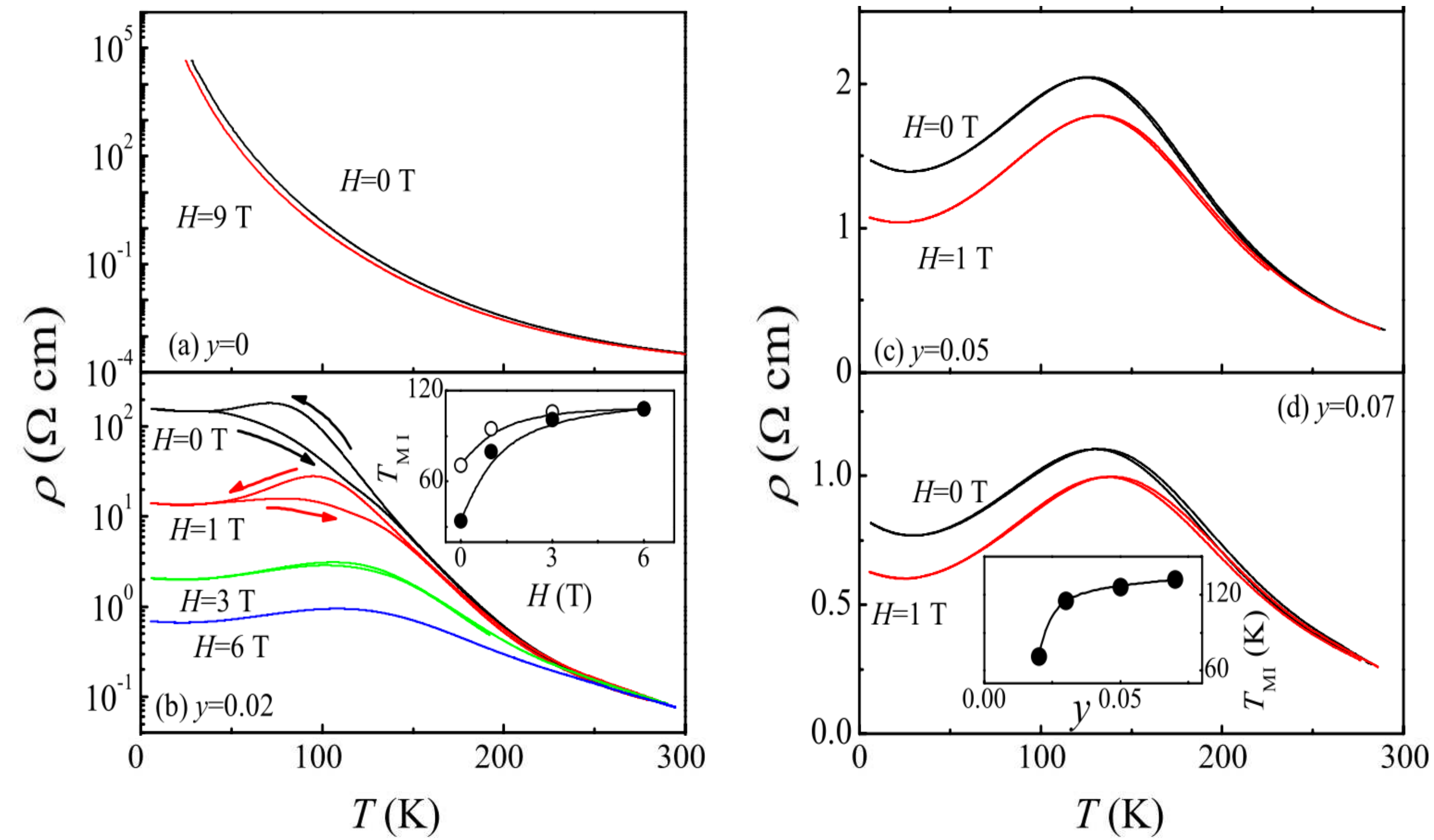}
\caption{Experimental spin resistivity versus $T$ are shown for several applied magnetic fields in the compound La$_{0.4}$Ca$_{0.6}$MnO$_{3}$. These data are in the figure 2 of Ref. \cite{Lu2009}.}
\label{Fig2_Lu}
\end{figure}

Figure \ref{J1J2_R_0.2_1.2} displays  the resistivity versus $T$ for $D_1=1.2$. The spin resistivity shows a rounded maximum at the transition temperature. This is in agreement with the curve experimentally observed in La$_{0.4}$Ca$_{0.6}$MnO$_3$ by Lu et al. \cite{Lu2009} (see Fig. \ref{Fig2_Lu}, right panel).

%Fig5

%Fig6

\begin{figure}[ht]
\centering
\includegraphics[width=10cm,angle=0]{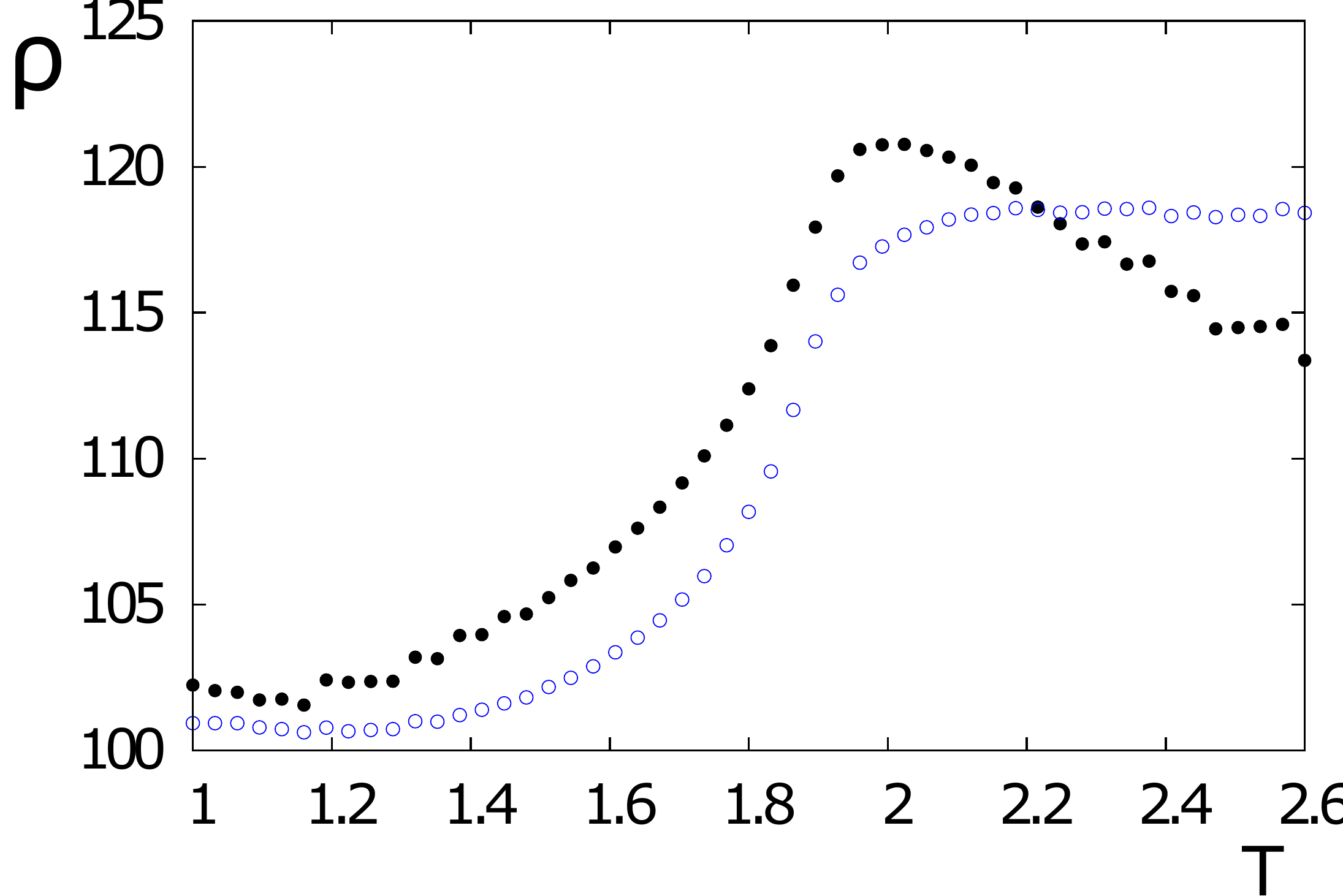}
\caption{$\rho$  as a function of $T$ for $\eta=0.2$ with $D_1=0.8$  and $1$ ((black circles and blue open circles, respectively). For simulations, we used  $N_x=N_y=20, N_z = 6$, $I_0=K_0=0.5$, $D_2=1$, $D=1$,  and $\epsilon=1$.}
\label{J1J2_R_0.2_0.8_1}
\end{figure}

For $D_1=0.8$ or $D_1=1$, the resistivity is smaller below the transition temperature, as seen in Fig. \ref{J1J2_R_0.2_0.8_1}.  This shows the importance of the effect of the interaction range on the spin resistivity in materials.

%Fig7

\begin{figure}[ht]
\vspace{0.5 cm}
\centering
\includegraphics[width=10cm,angle=0]{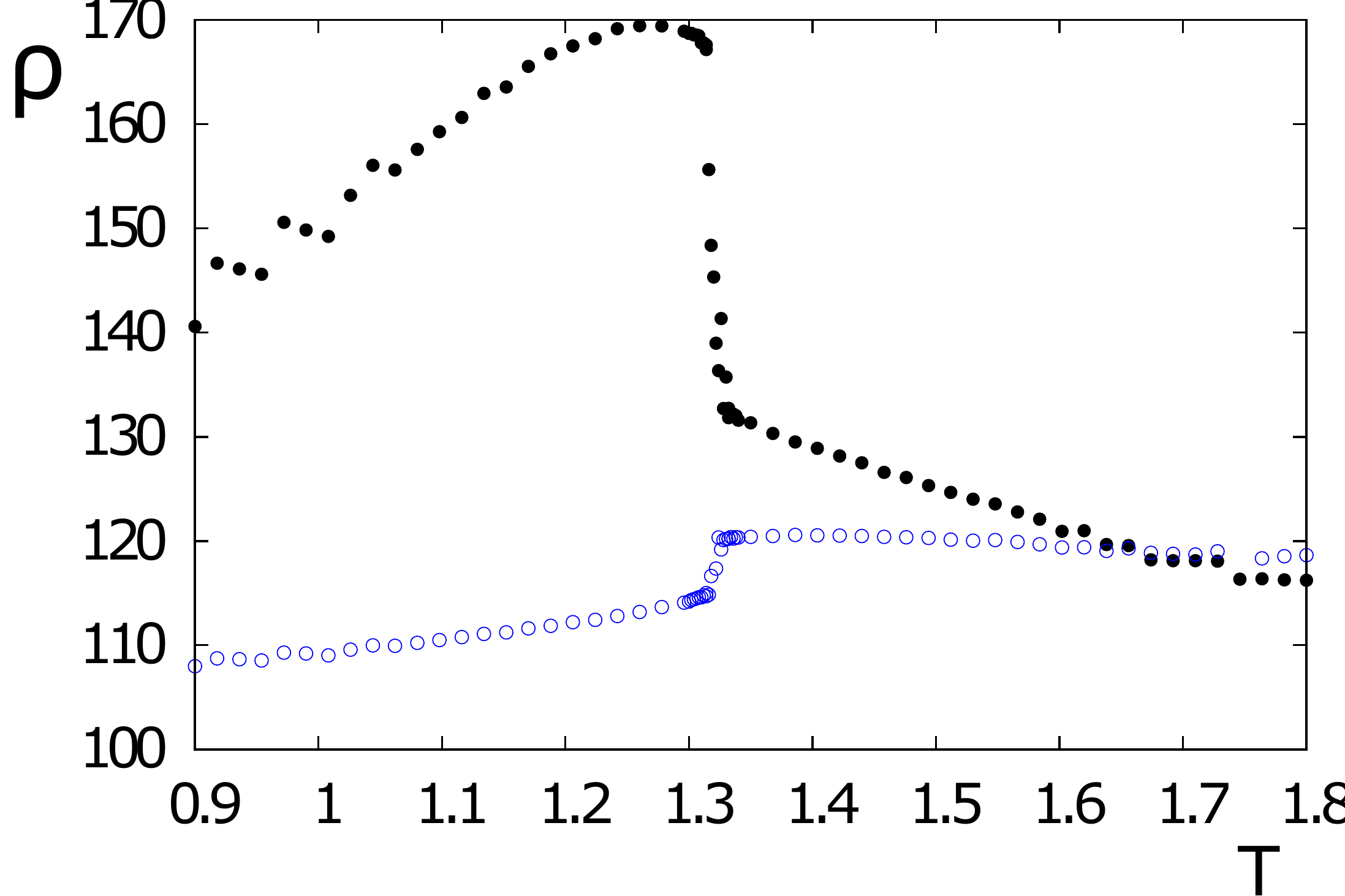}
\caption{ $\rho$ as a function of $T$  for $\eta=0.26$ where $D_1=0.8$ (black circles) and $D_1=1$ (blue open circles). We have used $N_x=N_y=20, N_z=6$, $I_0=K_0=0.5$, $D_2=1$, $D=1$  and $\epsilon=1$. See text for comments.}
\label{J1J2_R_0.26_0.8_1}
\end{figure}

\noindent \textbf{B. For $\eta=0.26$:}\\
At this value of $\eta$, the  transition of the lattice spins is of first order.   The dependence of the resistivity on  $D_1$ is very  similar to that of the second-phase transition, namely the resistivity at a given $T$ oscillates as $D_1$ varies. The physical meaning of the oscillation has been given above. More details can be found in Refs. \cite{Hoang2011,Magnin2}. 
We found that the resistivity $\rho$ in the frustrated regime can go downward or upward at the transition temperature  depending on  $D_1$ \cite{Hoang2011}, unlike in non-frustrated ferromagnets and antiferromagnnets shown earlier.  This is displayed in  Fig. \ref{J1J2_R_0.26_0.8_1} for two values of $D_1$ where one observes the discontinuity of $\rho$ at the transition temperature. The discontinuity of $\rho$  has also been found in other frustrated antiferromagnets such as the FCC antiferromagnet \cite{Magnin2}.

%Fig8

From the results shown above for the $J_1-J_2$ model, we come to the conclusion that the behavior of the spin resistivity is a consequence of the nature of the lattice transition. If the lattice transition is of second order, then the resistivity of itinerant spins has a rounded peak,  while if the lattice transition is of first order, the resistivity is  discontinuous at the transition temperature.
%%%%%%%%%%%%%%%%

%%%%%%%%%%%%%%%%

\subsection{The case of MnTe}\label{MnTe}
%Fig9
\begin{figure}[h!]
\vspace{1cm}
 \centering
 \includegraphics[width=60mm,angle=0]{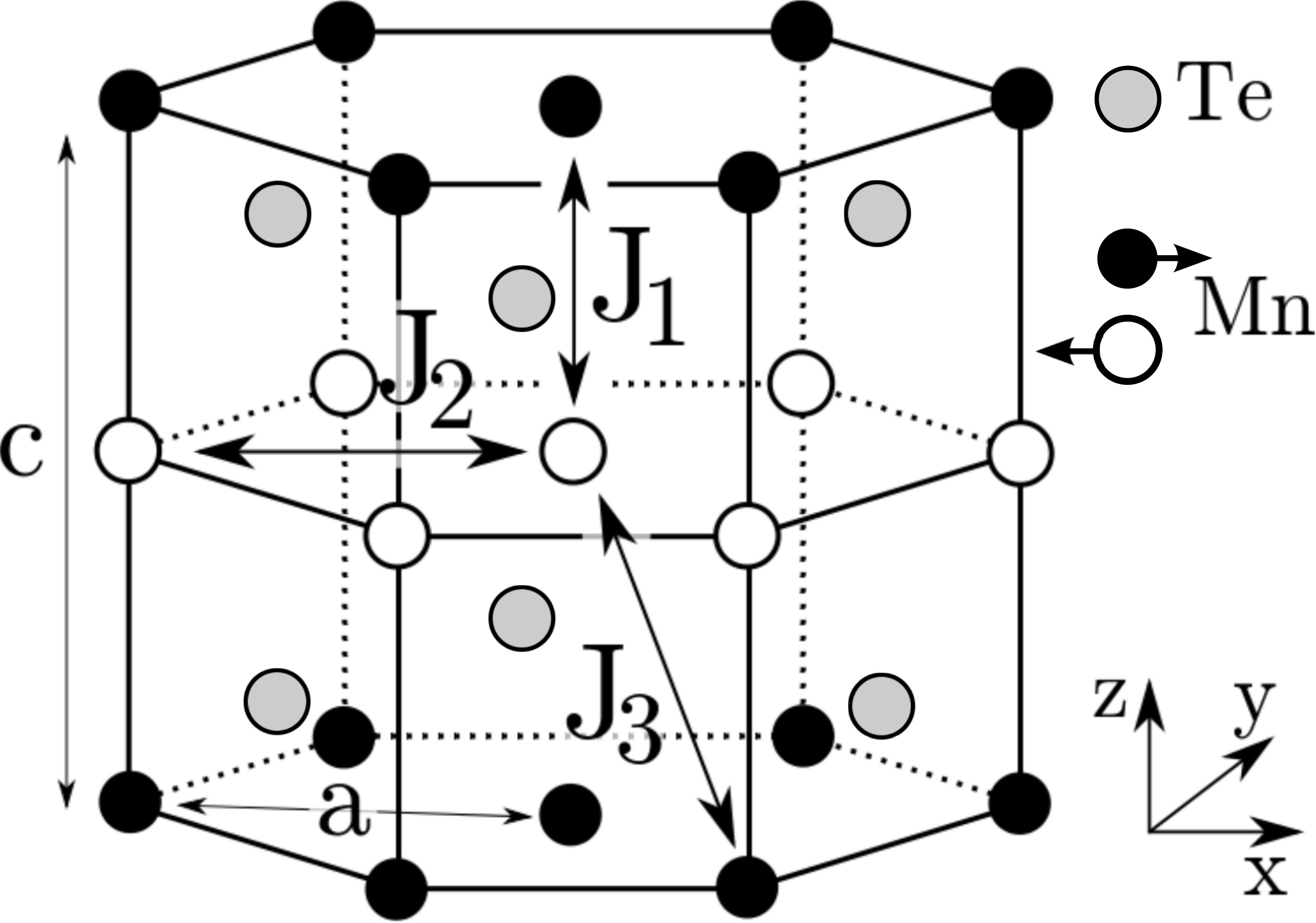}
\vspace{1cm}
 \caption{Structure of MnTe of  NiAs type: black and white circles present respectively opposite spins.  Interactions between NN, between next NN and between third NN are indicated respectively by $J_1$,  $J_2$, and  $J_3$.  See values given in the text.} \label{NiAs}
\end{figure}

The pure semiconductor MnTe has two kinds of structures:  the zinc-blend structure or the hexagonal NiAs one shown in Fig.~\ref{NiAs}  \cite{Hennion}.    We focus on the second structure where the N\'eel temperature is $T_N=310$ K \cite{Hennion2}, and where many other experimental data are available.
MnTe is a semiconductor with a large gap (1.27 eV) and a  carrier concentration 
$n=4.3\times 10^{17}$cm$^{-3}$ at room temperature \cite{Mobasser,Allen}.  Without doping, MnTe is non degenerate.
The crystal is formed by ferromagnetic $xy$ hexagonal planes antiferromagnetically stacked in the $c$ direction.  The NN distance in the $c$ direction is  $c/2\simeq 3.36$ \AA , and the in-plane NN distance is $a=4.158$ \AA.
From neutron scattering experiments,  it was found that the main exchange interactions between Mn spins in MnTe
are  the interaction between NN along the $c$ axis with the value  $J_1/k_B=-21.5\pm 0.3$ K,  the  ferromagnetic exchange
$J_2/k_B \approx 0.67\pm 0.05$ K between in-plane  neighboring Mn (they are next NN by distance),  and the  third NN antiferromagnetic interaction $J_3/k_B\simeq -2.87\pm 0.04$ K  (see Fig. \ref{NiAs}).
The spins are lying in the $xy$ planes perpendicular to the $c$ direction with a small in-plane easy-axis anisotropy $D_a$ \cite{Hennion2}.
Let us emphasize that  the values of the exchange integrals given above were deduced from experimental data by fitting with a free spin-wave theory \cite{Hennion2}. Other fittings with mean-field theories give slightly different values: $J_1/k_B=-16.7$ K, $J_2/k_B= 2.55$ K and $J_3/k_B= -0.28$ K  \cite{Mobasser}.  Note that   the Mn spin is experimentally known to be of
the Heisenberg model with  magnitude $S=5/2$ \cite{Hennion2}.

We write the following Hamiltonian for the lattice spins

\begin{eqnarray}
{\cal H} &=& -J_1\sum_{(i,j)} \vec{S}_i \cdot \vec{S}_j -J_2\sum_{(i,m)} \vec{S}_i\cdot \vec{S}_m
-J_3\sum_{(i,k)} \vec{S}_i \cdot \vec{S}_k  \nonumber \\
&& -D_a\sum_i (S_i^x)^2 \label{HLMnTe}
\end{eqnarray}
where the first sum is performed over the NN spin pairs, the second sum over the  NNN pairs and the third one over the third NN pairs.   $D_a>0$ is an anisotropy constant which favors the  in-plane $x$ easy-axis spin configuration.

The behavior of  $\rho$ in MnTe
 as a function of $T$ has been experimentally  shown in several works \cite{Chandra1996,Li,Russe,Efrem,He}.
We have studied using MC simulations the spin resistivity in MnTe with the above Hamiltonian \cite{Magnin2012}. Let us summarize this work here.

For MC simulations, we suppose the following Hamiltonian of the itinerant spins:

\begin{equation}\label{interact}
{\cal H}_i=-\sum_nI(\vec r-\vec R_n)\vec {\sigma} \cdot \vec S_n
\end{equation}
where the sum is performed by counting all  the lattice spins $\vec S_n$ inside the sphere of radius $D_1=a$ centered at  $\vec r$.  $I(\vec r-\vec R_n)>0$ is the ferromagnetic distance-dependent interaction between  the itinerant electron spin $\vec {\sigma}$ at $\vec r$ and the Mn spin $\vec S_n$ at $\vec R_n$ .

The electron spin is supposed of the Ising type. We neglect therefore the quantum effects which may be important at very low $T$ but our attention is focused on the region of high-enough $T$ where quantum effects may be neglected.
We assume the following form of $I(\vec r-\vec R_n)$ :

\begin{equation}\label{interact1}
I(\vec r-\vec R_n)=I_0\exp [-\alpha(\vec r-\vec R_n)]
\end{equation}
where the constants $I_0$ and $\alpha$ are chosen in such a way that the interaction ${\cal H}_i$ yields an energy much smaller than the lattice energy given by ${\cal H}$ (see the guide for the choice of different constants given below Eq. (\ref{K}) and in Ref. \cite{Magnin}).  It is noted  that the cut-off distance $D_1$ is rather short so that only the first few neighbors are inside the sphere, the results shown below do not therefore depend significantly on the choice of the value of $\alpha$ in the exponential. Finally, note that  the  concentration of conduction electrons in MnTe is
 $n=4.3\times 10^{17}$cm$^{-3}$ which is five orders lower than the concentration of its surrounding lattice spins which is
$\simeq 10^{22}$cm$^{-3}$.   This observation justifies that the interaction between conduction electrons for MnTe can be neglected. We have assumed this in the simulations shown in the following.

As mentioned above,  the exchange interactions deduced from experimental  data have slightly different values, they depend on the theoretical Hamiltonian and the approximations  used to deduce it (often the mean-field approximation is used, see a detailed example in Ref. \cite{samia}).  Note that in semiconductors, the carrier concentration varies with  $T$ but since this concentration is very low, we do not take into account its variation.
Consequently, the number of conduction electron spins used in the simulation is important only for the statistical average.  The current obtained is proportional to the number of itinerant spins but there are no extra effects within our assumptions mentioned above.  

\begin{figure}[h!]
\vspace{1cm}
 \centering
 \includegraphics[width=100mm,angle=0]{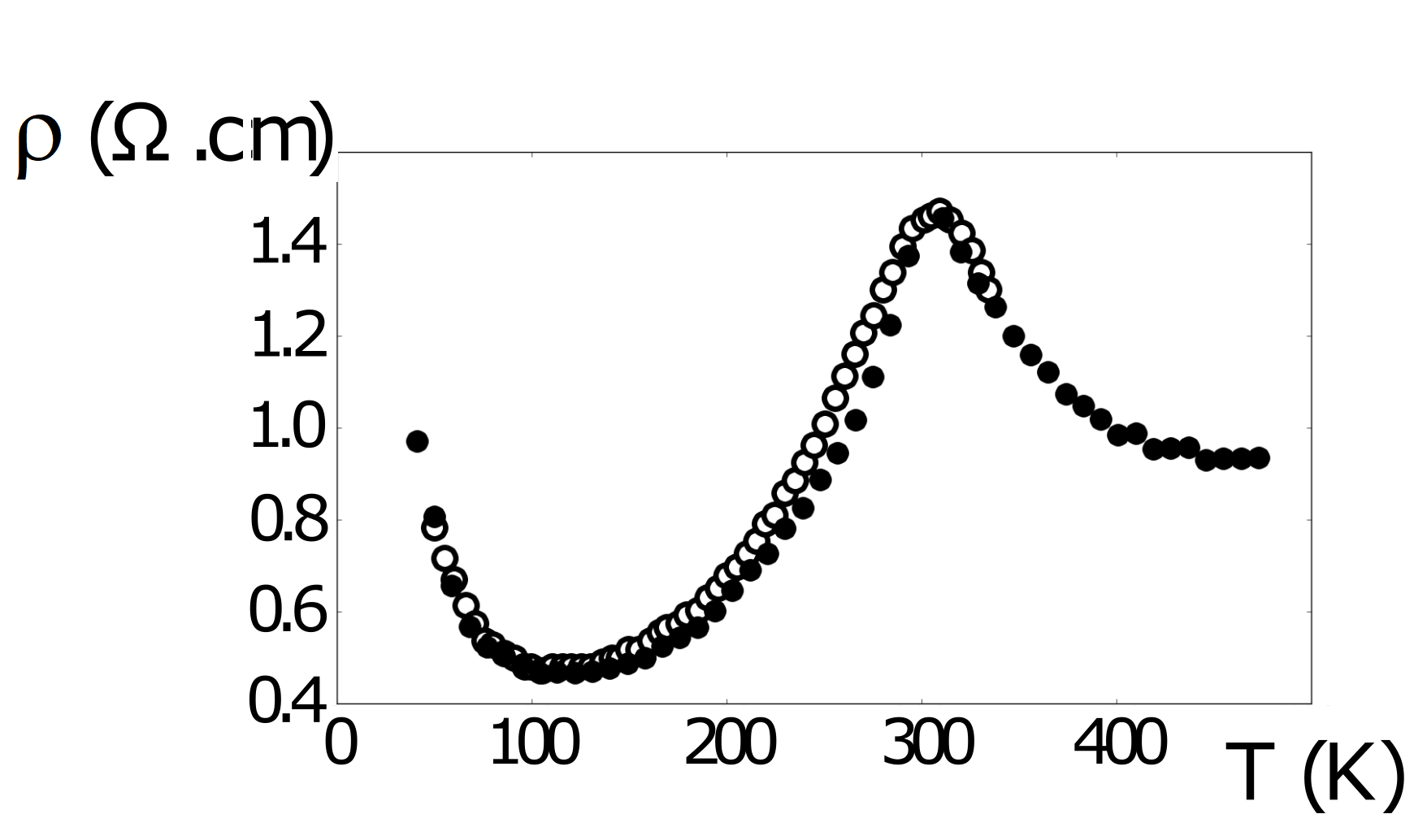}
 \caption{Comparison between the simulated spin resistivity and the experimental data of MnTe:   Black circles are results from the Monte Carlo simulation, white circles are experimental data taken from He et al.\cite{He}. We used for the simulation $J_1=-21.5$K, $J_2=2.55$ K, $J_3=-9$ K, $
I_0=2$ K, $D_a=0.12$ K, $D_1=a=4.148$ \AA, $\epsilon = 2\times10^5$ V/m, $L=30a$ (lattice size: $L^3$). See text for comments.} \label{RMnTe}
\vspace{1cm}
\end{figure}

We have calculated $\rho$ of  MnTe, using the exchange integrals slightly modified with respect to the ones given above in order to obtain the best fit. 
The obtained resistivity $\rho$ is shown in
Fig.~\ref{RMnTe}. Let us note that we have taken $J_3$ slightly larger in magnitude than the value deduced from experiments by mean-field approximation. Our value of $J_3$ was chosen in order to obtain $T_N=310$ K
which is in excellent agreement with experiments. However the most striking feature is that the simulated  $\rho$ shows a sharp maximum at $T_N$ and coincides with the experimental data over the whole temperature range. Note that we have used $A=1$ and the well-known Heisenberg critical exponents  $\nu=0.707$, $z=1.97$ \cite{Peczak} for the lattice spins.  It is remarkable that with the same set of param:eters, we obtain an excellent agreement with experiments in the temperature regions below $T<140$ K and above $T_N$. We note that we have tried earlier to use the Boltzmann's equation \cite{Akabli4} but the obtained result is not as good as the MC result presented above.

From the simulated $\rho$, we can calculate the relaxation time  of conduction spins, we obtain  $\tau_I \simeq 0.1$ ps. The mean free path can be also estimated,  it is equal to $\bar l \simeq  20 $ \AA, at the critical temperature.  

%Fig10

%%%%%%%%%%%%%%

\section{Phase transition and spin resistivity in the Ising HCP lattice}\label{hcplattice}
\subsection{Hamiltonian and Ground State}\label{modelhcp}

\begin{figure}
\centering
\includegraphics[width=50mm,angle=0]{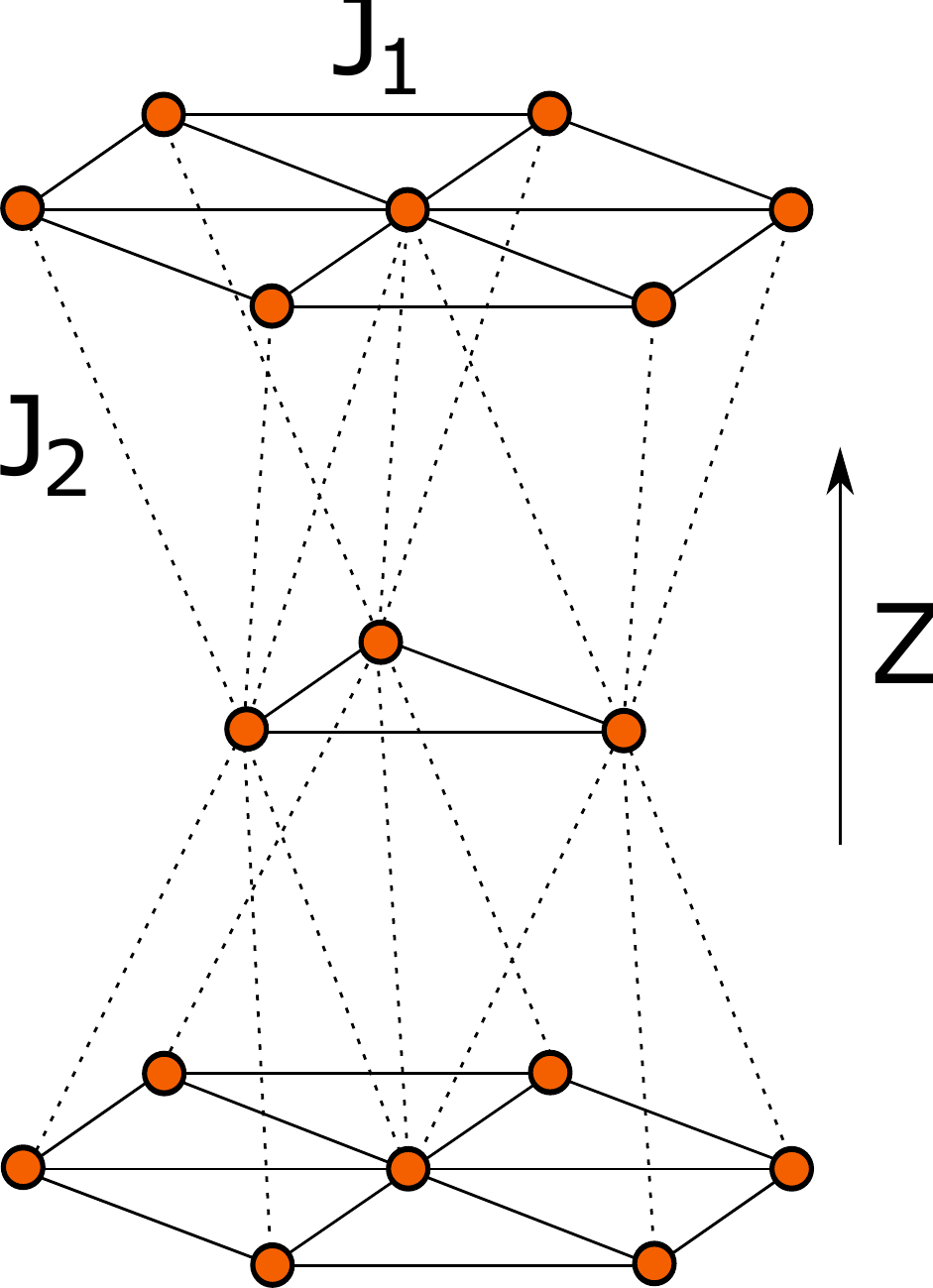}
\caption{HCP lattice: the in-plane NN interaction is denoted by $J_1$ and the inter-plane NN interaction is denoted by $J_2$.} \label{hcp}
\end{figure}

The lattice we consider is the HCP structure illustrated in Fig. \ref{hcp}. The $xy$ planes are triangular (hexagonal) and the stacking direction is $z$.  We suppose the following  Hamiltonian 

%Fig11

\begin{equation}\label{HLHCP}
{\cal H} = -\sum_{(i,j)}J_{i,j}\vec S_{i} \cdot \vec S_{j}
\end{equation}
where $J_{ij}$ is the AF interaction between nearest-neighbors (NN) $\vec S_{i}$ and $\vec S_{j}$. We denote $J_{ij}=J_1$ if the NN are on the $xy$ triangular plane, and $J_{ij}=J_2$ if the NN are on two adjacent planes (see Fig. \ref{hcp}).    The GS can be determined by minimization the local energy of each spin and doing this for all spins,  then repeating many times until the total energy converges to a minimum. Normally, with a system without bond disordering, this method needs just a small number of iterations. The GS can be checked by looking at the final snapshot: it should be periodic.   This procedure of local energy minimization is called in the literature "the steepest-descent method".
The implementation of this method is very simple \cite{Ngo2007} (i) we first create an initial random configuration (ii) we then calculate the local field acting at a spin by its neighbors using  (\ref{HLHCP})  (iii) we align the spin under consideration along the calculated local field, in doing so its energy is minimum  (iv) we take another spin and repeat the three preceding steps until all spins are considered: this step completes one sweep (v) we  start again another sweep and we realize a large number of sweeps until the total energy is minimm.

One can also analytically minimize the interaction energy as shown below to find the GS.  Let  us assume that both interactions $J_1$ and $J_2$ are antiferromagnetic. For simplicity we fix $J_2=-1$ and vary $J_1$.

The case of isotropic interaction, namely $J_1=J_2$ has been studied in Ref. \cite{Diep1992}. We summarize the result here: for the HCP structure, each spin is common for eight tetrahedra (four in the upper half-space and four in
the lower half-space along the $z$ axis) and a NN bond is
shared by two tetrahedra. The GS spin configuration of
the system is formed by stacking neighboring tetrahedra.
In the GS, one has two pairs of antiparallel spins on each tetrahedron.  Their axes form an arbitrary angle
 $\alpha$.  The GS degeneracy is therefore infinite (see Fig. 2a of Ref. \cite{Diep1992}). Note  tthat the periodic boundary conditions
will reduce a number of the GS configurations, but the degeneracy is still infinite. One particular
family of configurations of interest for both XY and Heisenberg cases is when $\alpha=0$. The GS is then collinear with two spins up and the other two down. The stacking sequence is simple because there are three equivalent configurations due to the fact that there are three ways
to choose the parallel spin pair in the original tetrahedron.

The case where $J_1\neq J_2$ has been studied in  Ref. \cite{Hoang2012} for the Ising and XY cases. Let us recall some results concerning the Ising case which allow us to understand the new results on the spin resistivity presented below.

%\begin{itemize}
%\item Ising case:

We use the steepest descent method described above with varying $J_1$ ($J_2=-1$): we find two kinds of GS spin configuration: the first consists of $xy$ ferromagnetic planes stacked antiferromagnetically along the $z$ direction, while the second one is the stacking of $xy$ AF planes such that each tetrahedron has two up and two down spins.  The transition between the two configurations is determined as follows: one simply writes down the respective energies of a tetrahedron and compares them

\begin{eqnarray}
E_1&=&3(-J_1+J_2)\\
E_2&=&J_1+J_2
\end{eqnarray}
One sees that $ E_1<E_2$ when $J_1>0.5J_2$,  i.e.  $|J_1|<0.5 |J_2| $.
Thus the first configuration is more stable when $|J_1|<0.5 |J_2| $.

\subsection{Phase Transition in the case of Ising Spins on the HCP Lattice}

 In the following, we present the results of simulations using the Hamiltonian Eq. (\ref{HLHCP}). We use the sample size  $N_x \times N_y \times N_z$ with $N_x = N_y = 18$ and $N_z=8$, namely 16 atomic planes along the $z$ axis, and the periodic boundary conditions in all directions. We use the first $10^6$ MC steps per spin to reach equilibrium and we average physical quantities with the next $10^6$ MC steps per spin. The energy is expressed in the unit of $|J_2|=1$.

%\noindent {\bf A. Ising case}

\begin{figure}[ht]
\centering
\includegraphics[width=5.5cm,angle=0]{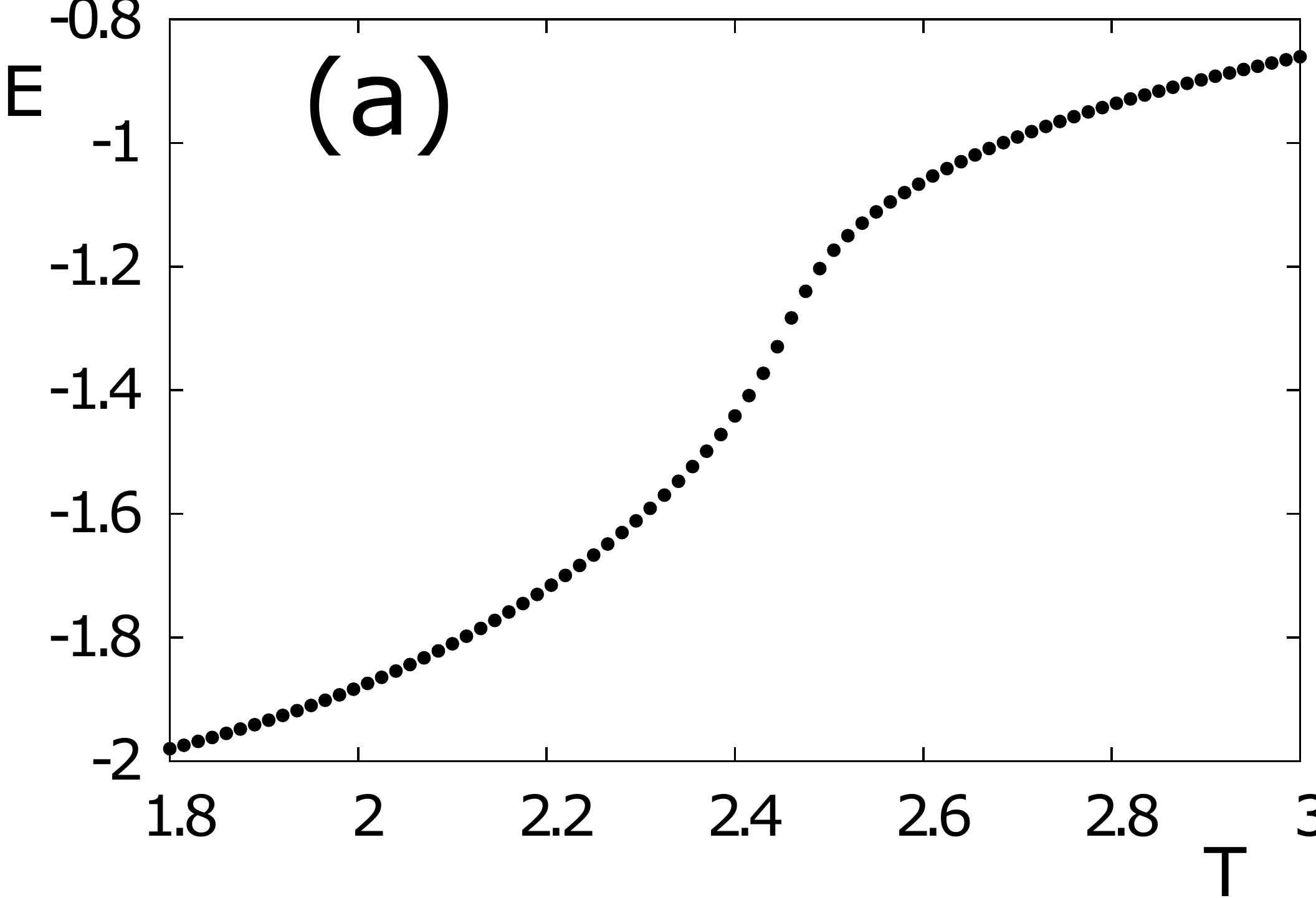}
\includegraphics[width=5.5cm,angle=0]{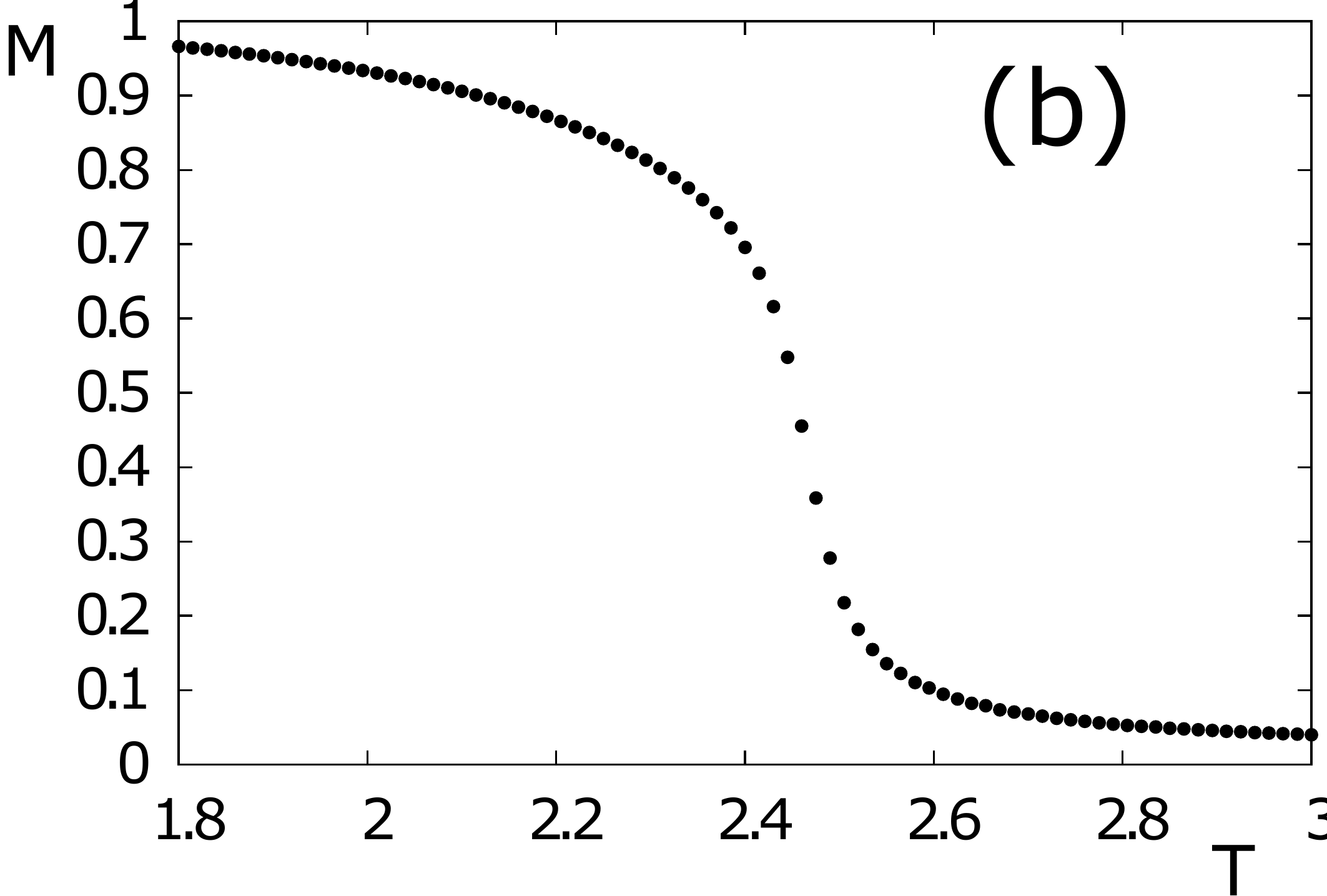}
\includegraphics[width=5.5cm,angle=0]{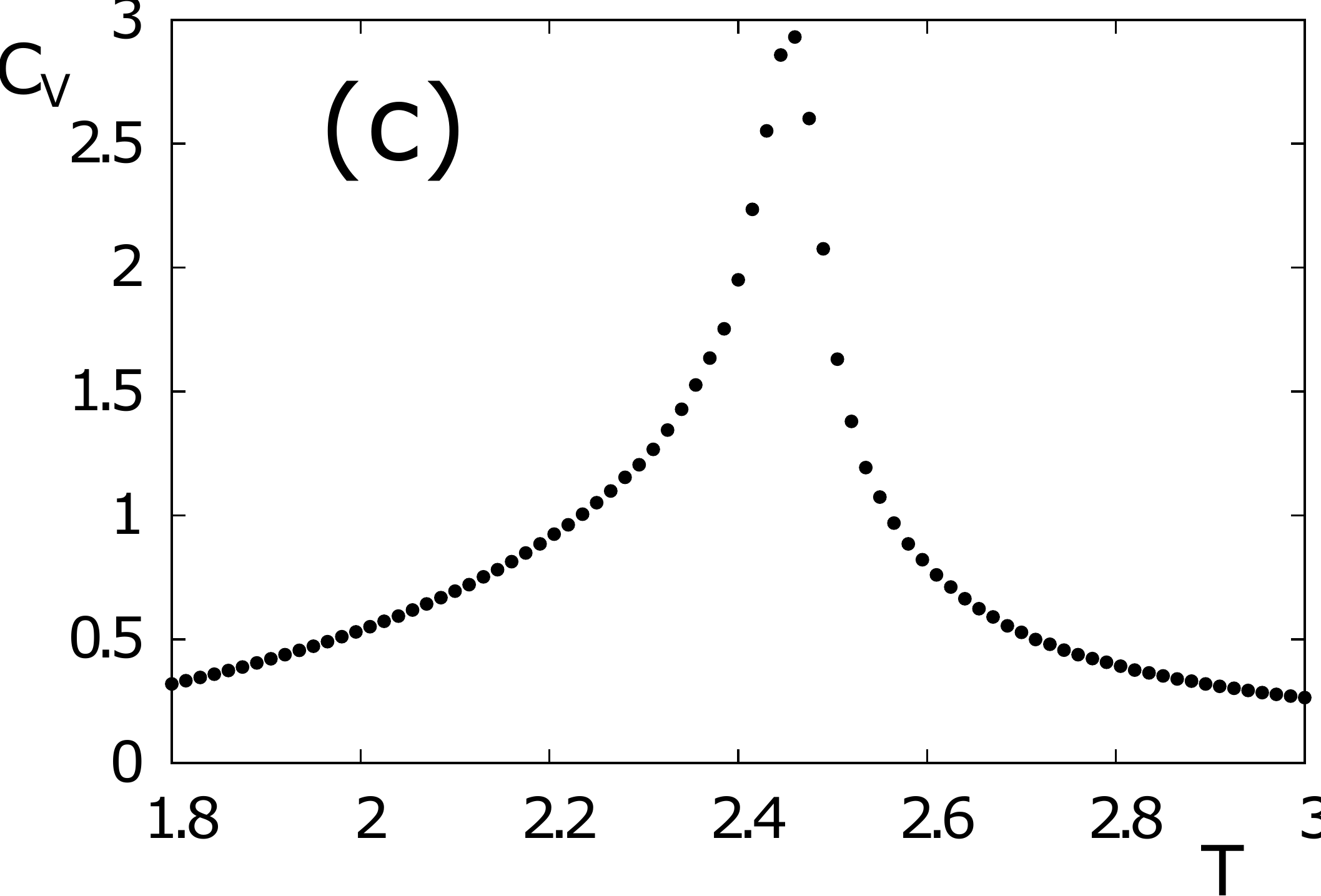}
\includegraphics[width=5.5cm,angle=0]{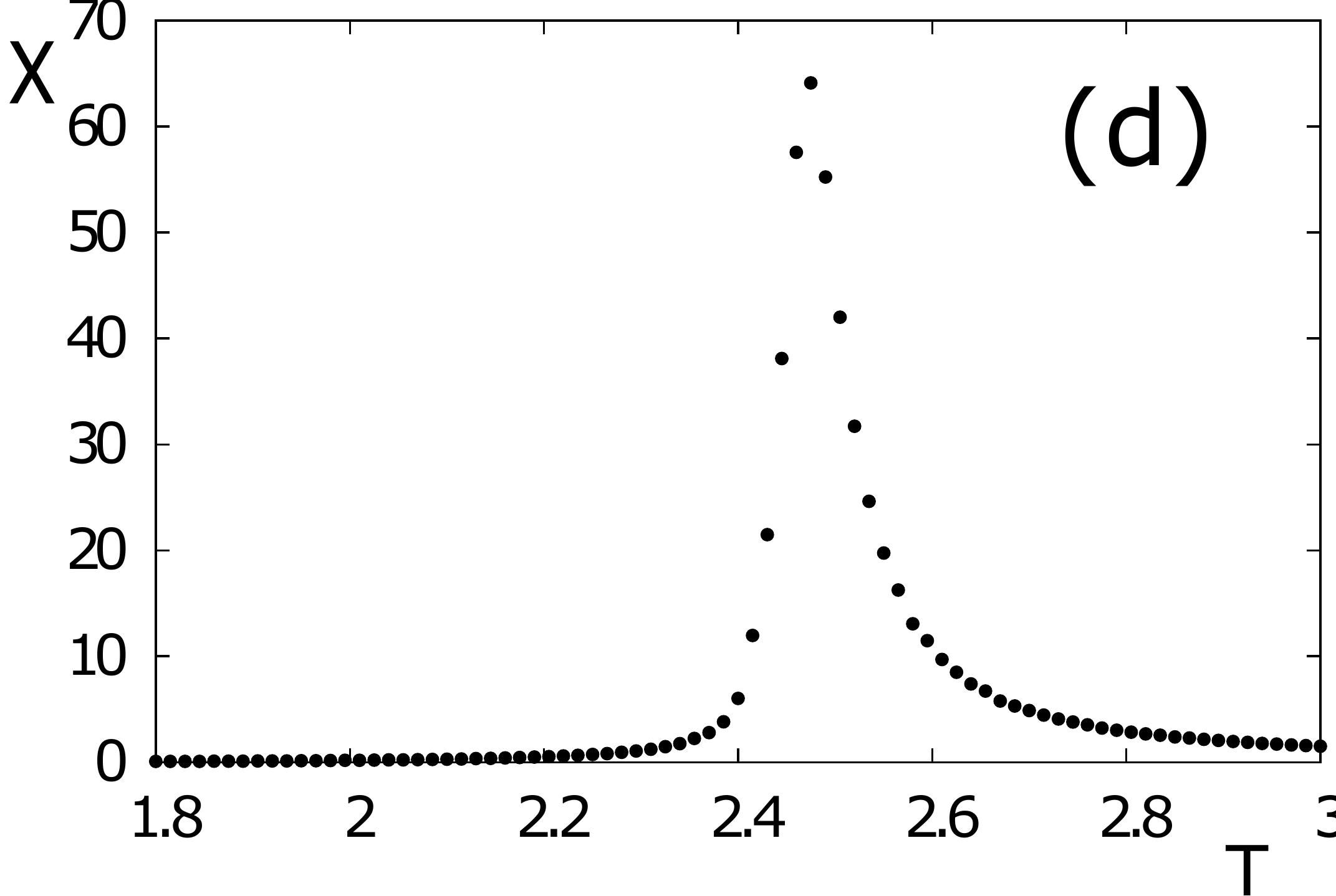}
\caption{The case of Ising spin  on the antiferromagnetic HCP lattice: (a) energy per spin $E$, (b) order parameter $M$,  (c) specific heat $C_V$ and (d) susceptibility $\chi$,  versus temperature $T$ for $\eta=J_1/J_2=0.30$. See text for comments.}
\label{HCP_0.3_E}
\end{figure}

Let us define $\eta=J_1/J_2$.  We have seen that the GS changes at $\eta_c=0.5$, so we show below the properties of the system on both sides of this value. Figure \ref{HCP_0.3_E} displays the averaged energy per spin, the order parameter (staggered magnetization), the specific heat and the susceptibility for $\eta=0.3$. As seen, the transition is of second order since there is no discontinuity of the energy and the order poarameter at the transition temperature.

%Fig12

\begin{figure}[ht]
\vspace{0.5cm}
\centering
\includegraphics[width=5.5cm,angle=0]{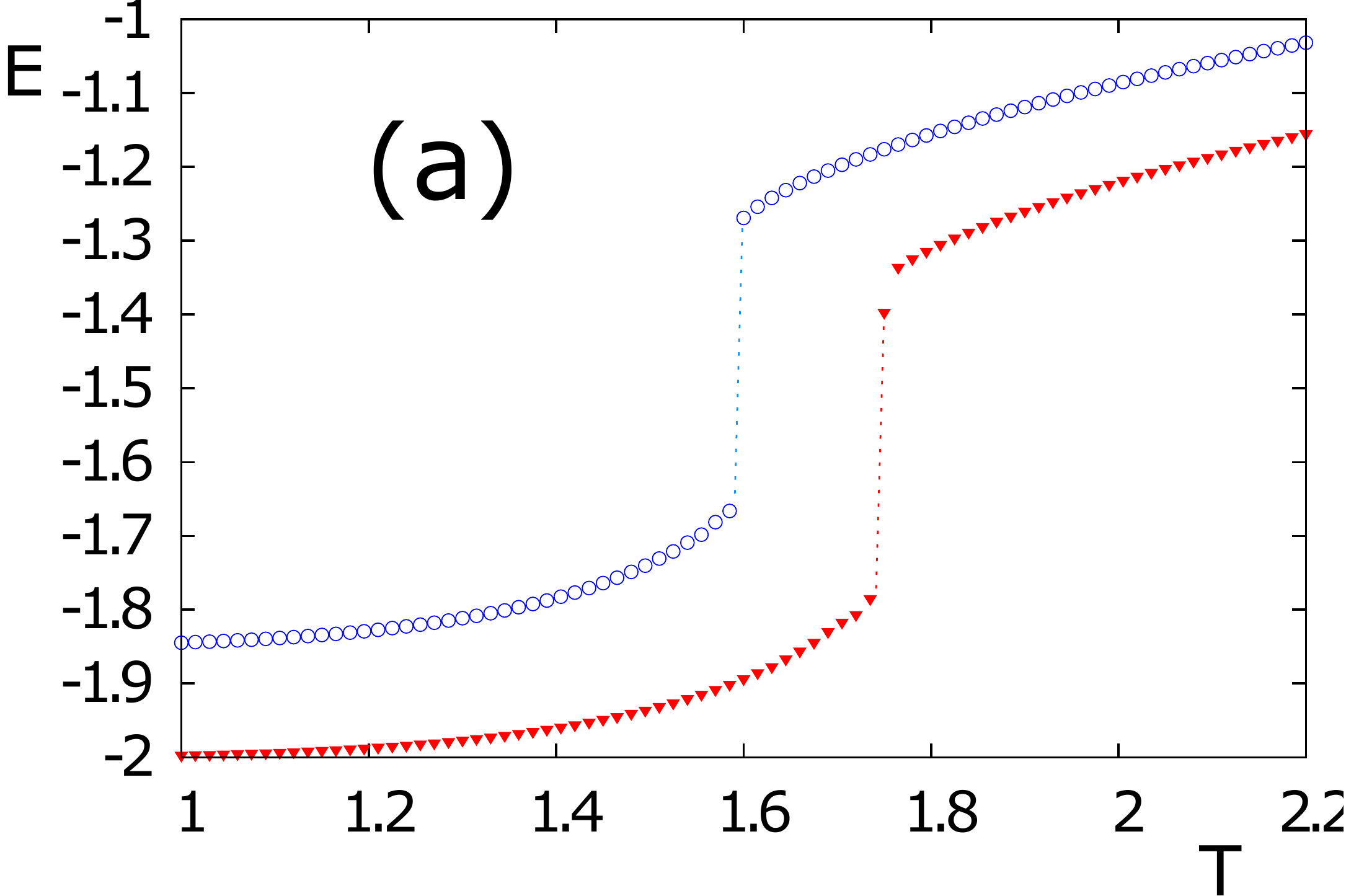}
\includegraphics[width=5.5cm,angle=0]{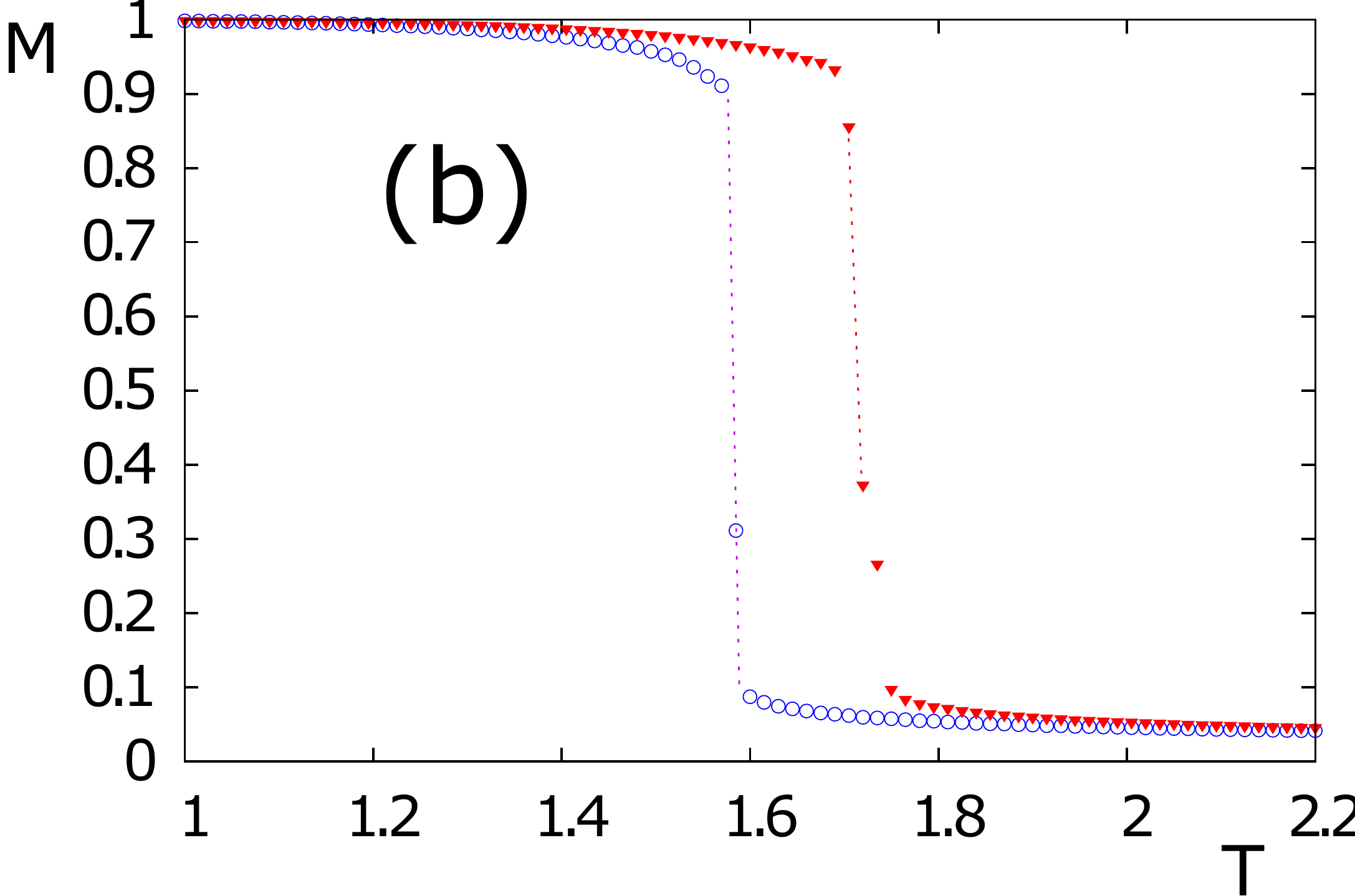}
\caption{The case of Ising spin on the antiferromagnetic HCP lattice: (a) energy per spin $E$; (b) order parameter $M$ versus temperature $T$ for $\eta=J_1/J_2=0.85$ (blue open circles) and $1$ (red triangles). See text for comments.}
\label{HCP_0.85_1_E}
\end{figure}

For $\eta=J_1/J_2>0.5$,  Fig. \ref{HCP_0.85_1_E} for   $\eta=0.85$ and $1$ shows that  the discontinuity of $E$ and $M$ at the transition is very large,
a signature of a strong first-order transition in both cases.
%Fig13

\begin{figure}[ht]
\centering
\vspace{0.5cm}
\includegraphics[width=8cm,angle=0]{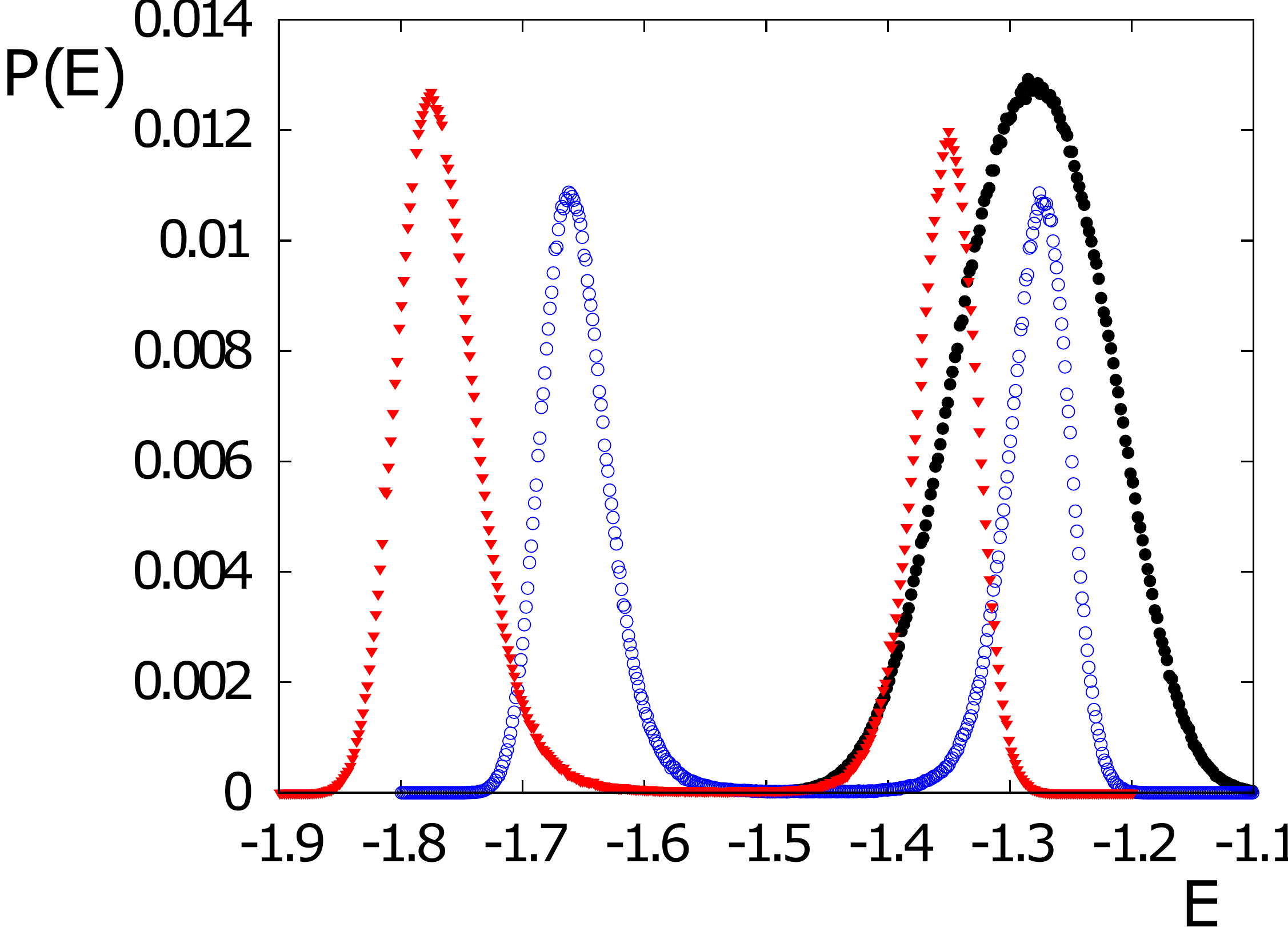}
\caption{Energy histogram $P(E)$  for $\eta=0.3$ (black circles), $0.85$ (blue open circles), and $1$(red triangles). See text for comments.}
\label{HCP_P}
\end{figure}

In order to confirm the order of the phase transition, we measure the energy histogram taken during the averaging MC time. Figure \ref{HCP_P} shows the energy histogram taken at the transition temperature for $\eta=0.3$ (black), $0.85$ (blue) and $1$ (red). We observe here that the first case is a Gaussian distribution indicating a second-order transition, in contrast to the last two cases which show  double-peak histograms confirming a first-order transition.  

%Fig14

\begin{figure}[ht]
\vspace{0.5cm}
\centering
\includegraphics[width=8cm,angle=0]{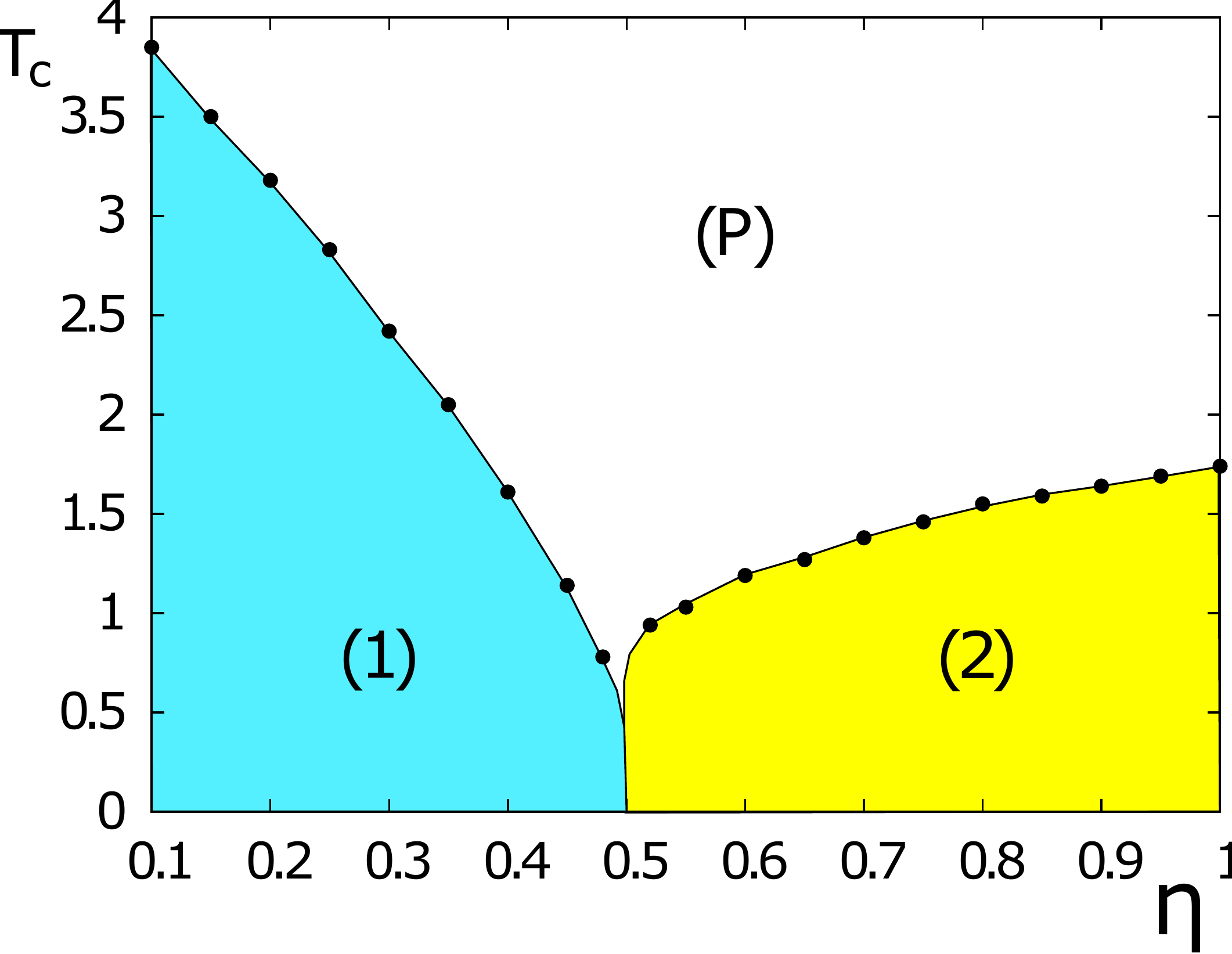}
\caption{Transition temperature $T_C$ versus $\eta$. (1) denotes the second-order region, (2) the first-order region and (P) the paramagnetic phase. }
\label{HCP_phase_diagram}
\end{figure}

Figure \ref{HCP_phase_diagram} displays the phase diagram in the space $(T_C,\eta)$ where zone (1) and zone (2) denote the ordering of the first, and second kinds, respectively; (P) indicates the paramagnetic phase. Note that the transition line between (1) and (P) is a second-order line, while that between (2) and (P) is a first-order line.

%Fig15

%\clearpage
%\noindent {\bf B. XY case}

Note that in the $XY$ case, the change of the GS takes place at $\eta_c=1/3$. We have studied this case in details in to Ref. \cite{Hoang2012}. 
%We show in Fig. \ref{HCP_XY_0.3_E} the result for $\eta=0.3$ where the GS is composed of ferromagnetic planes antiferromagnetically stacked in the $z$ direction. The transition is of second order.

%Fig18

Finally, let us emphasize that all 3D frustrated systems we know so far undergo a first-order transition \cite{DiepFSS} including the much-studied antiferromagnetic stacked triangular lattice \cite{Itakura2003,Bekhechi2006,Ngo2008a,Ngo2008b,Delamotte2005}, the FCC antiferromagnets \cite{Diep-Kawamura}, the simple cubic fully frustrated lattices \cite{Blankschtein1984,Ngo2010,Ngo2011a,Ngo2011b}, helimagnets \cite{Diep1989}, and antiferromagnetic HCP lattice studied here (see more details in Refs. \cite{Diep1992,Hoang2012}).

\subsection{Spin resistivity in the HCP lattice with Ising spins}

\begin{figure}[ht]
\centering
\includegraphics[width=8cm,angle=0]{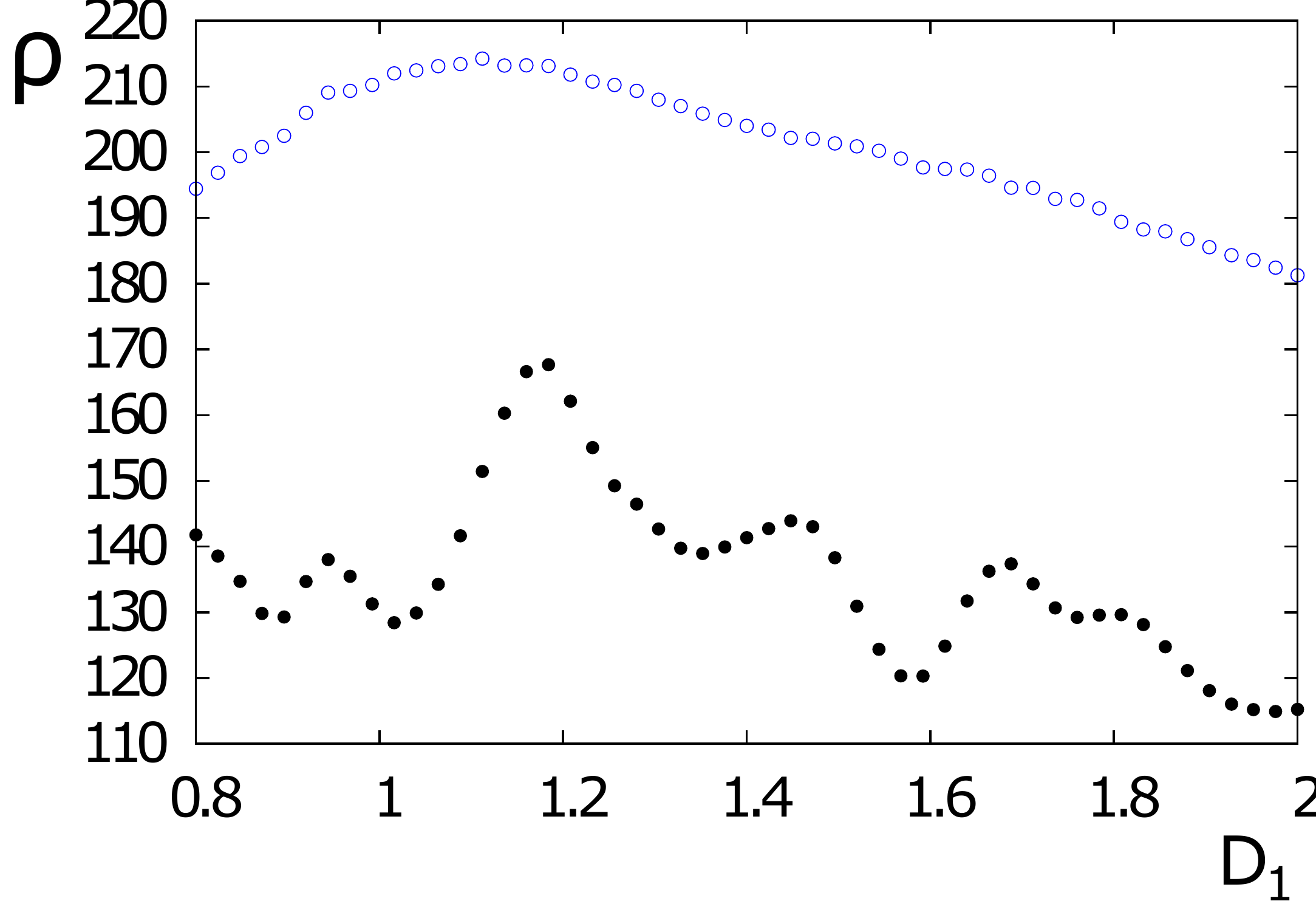}
\caption{$\rho$ versus $D_1$ for $\eta=0.3$ at $T=1.6<T_C$ (black circles) and at $T=2.8>T_C$ (open circles) where $T_C\simeq 2.4$. Other parameters are $N_x=N_y=18$, $N_z=8$, $D_2=1$, $I_0=2$, $K_0=0.5$, $C_1=C_2=1$, $A=1$, $D=0.5$, $\epsilon=1$.}
\label{HCP_R_effectD1_0.3}
\end{figure}

\begin{figure}[ht]
\vspace{0.5cm}
\centering
\includegraphics[width=8cm,angle=0]{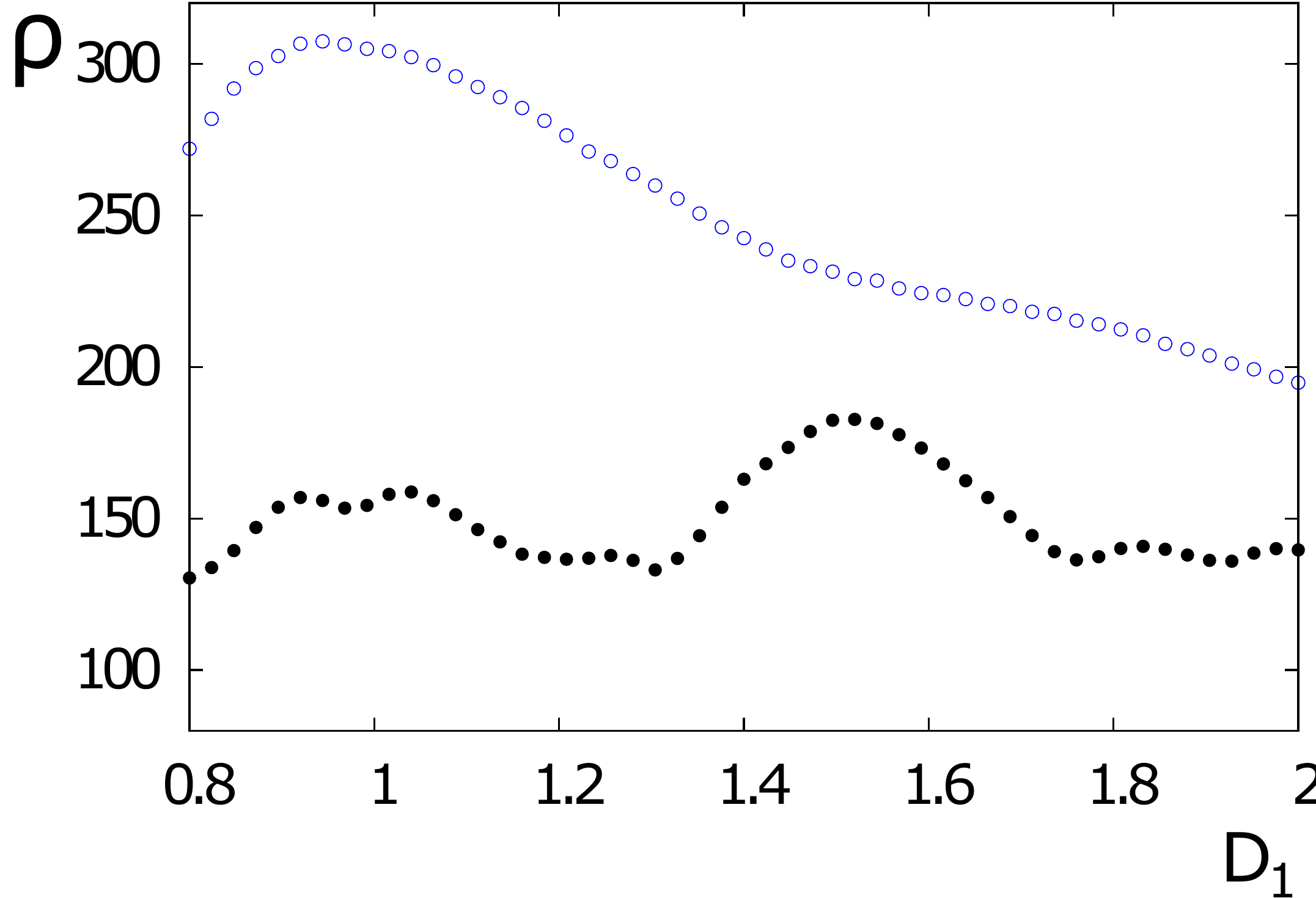}
\caption{$\rho$ versus $D_1$ for $\eta=1$ at $T=1.5<T_C$ (black circles) and at $T=1.9>T_C$ (open circles) where $T_C\simeq 1.7$. Other parameters are $N_x=N_y=18$, $N_z=8$, $D_2=1$, $I_0=2$, $K_0=0.5$, $C_1=C_2=1$, $A=1$, $D=0.5$, $\epsilon=1$.}
\label{HCP_R_effectD1_1}
\end{figure}

The results in this subsection are new, they are not published so far. Using the method which has been described in subsection \ref{method}, we carry out MC simulations to study the spin resistivity in the Ising case. We show in Fig. \ref{HCP_R_effectD1_0.3} the resistivity at two temperatures, below and above the transition temperature, as a function of $D_1$ for the GS belonging to phase (1). We show in Fig. \ref{HCP_R_effectD1_1}  the case of a GS belonging to  phase (2) (see Fig.  \ref{HCP_phase_diagram}).  Similar to the case of $J_1-J_2$ model on the simple cubic lattice considered in Ref. \cite{Hoang2011,Magnin2}, we find here an oscillation of $\rho$ at low temperature. Note that  $\rho$ is always smaller at low temperature than at high temperature, whatever the value of $D_1$ is. The physical origin of the oscillation has been discussed above. 

%Fig16

%Fig17

\begin{figure}[ht]
\centering
\includegraphics[width=6cm,angle=0]{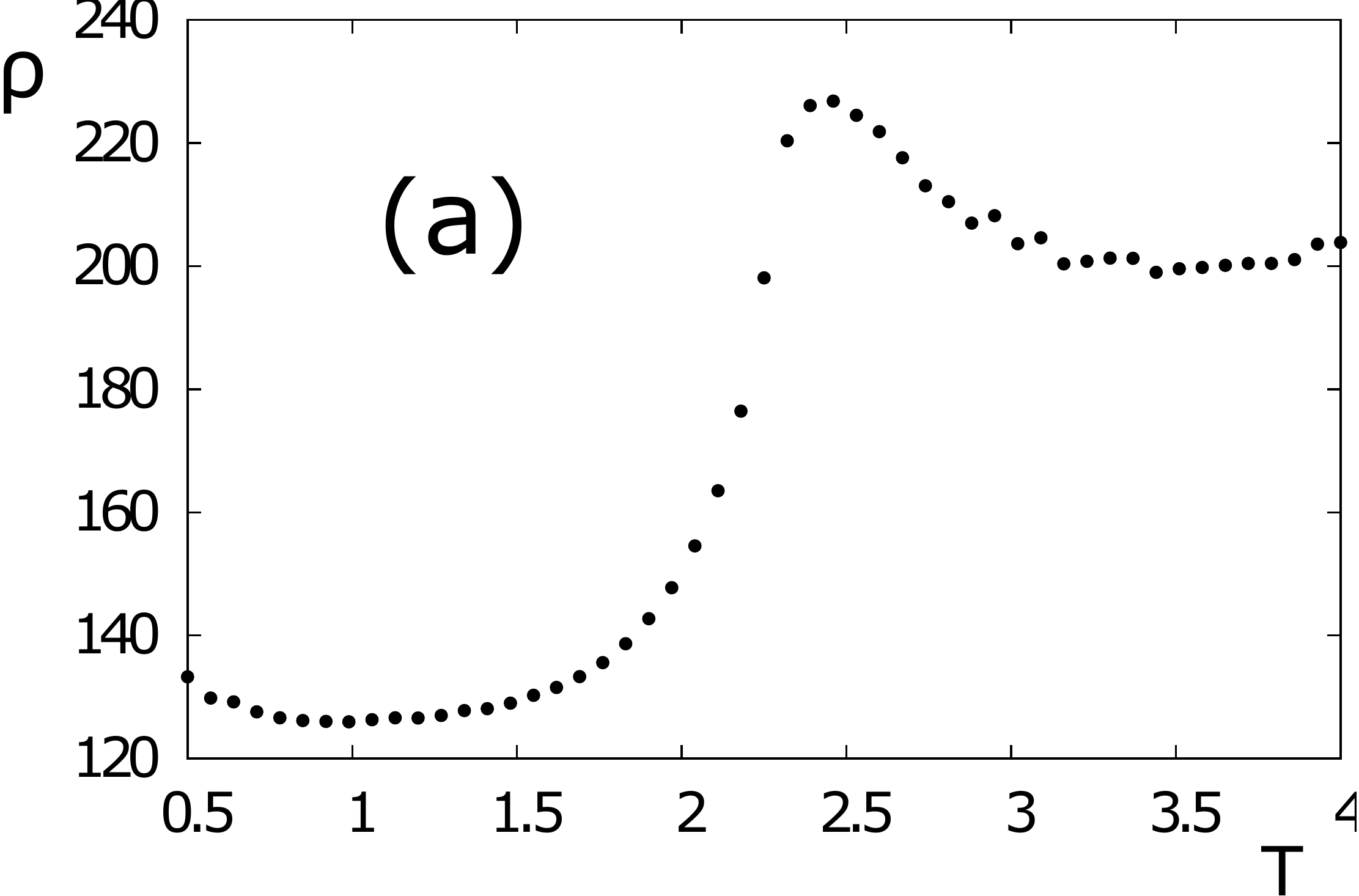}
\includegraphics[width=6cm,angle=0]{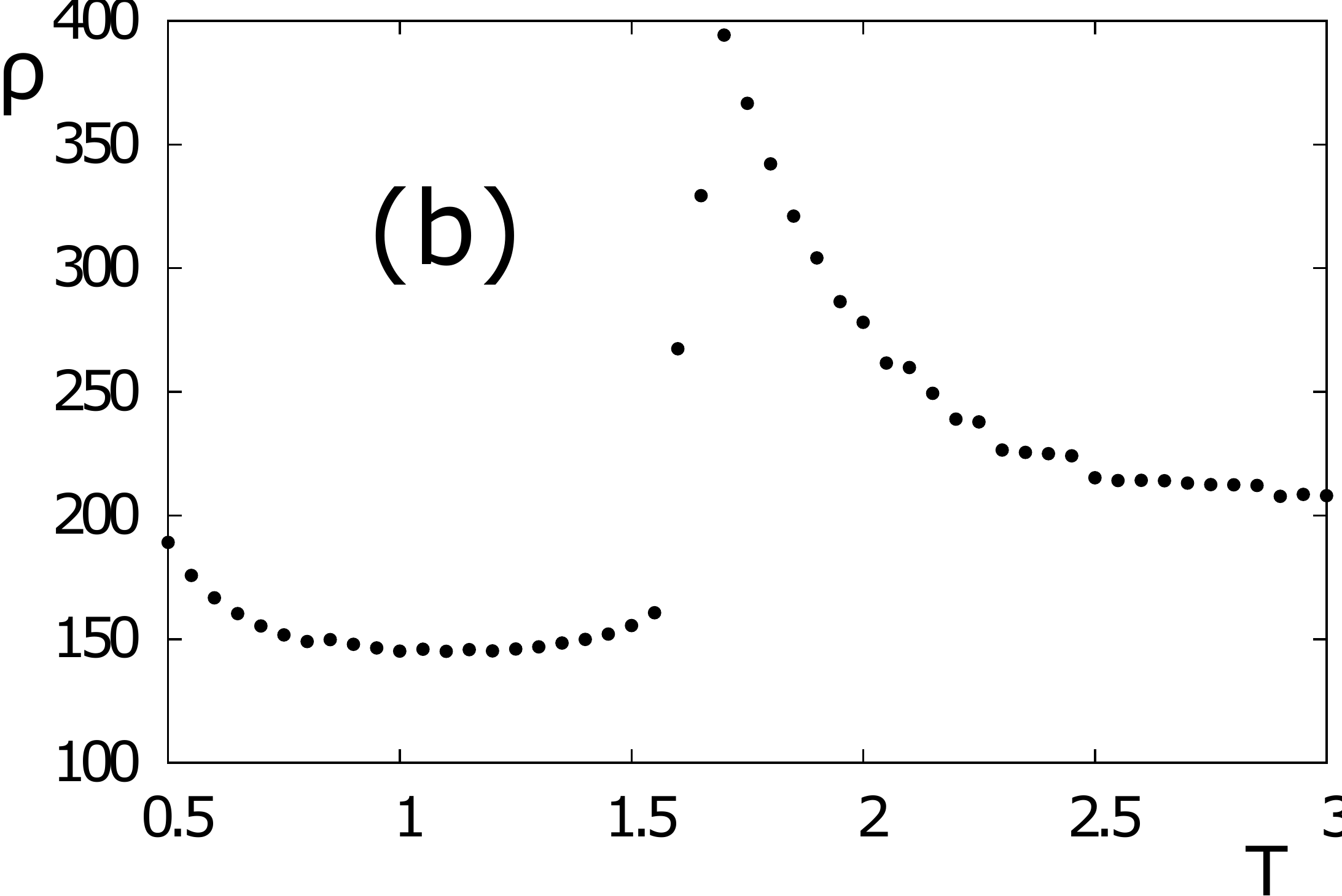}
\caption{Spin resistivity $\rho$ of the Ising HCP model versus temperature $T$ for (a) $\eta=0.3$;  (b) $\eta=1$. $N_x=N_y=18$, $N_z=8$ (namely $16$ planes in the $z$ direction), $D_1=D_2=1$, $\epsilon=1$, $I_0=2$, $K_0=0.5$, $C_1=C_2=1$, $A=1$, $D=0.5$. All distances are in unit of the NN distance, energy constants are in unit of $|J_2|=1$.  See text for comments.} \label{HCP_R}
\end{figure}

The spin resistivity $\rho$  for  $\eta=0.3$ and $1$ is shown in Fig. \ref{HCP_R} as a function of $T$, here the distances $D_1$ and $D_2$ are in unit of the distance between the NN lattice spins, and  $I_0$, $K_0$ and $D$ which have the energy dimension are in the unit of $|J_2|=1$. As in the frustrated $J_1-J_2$ model shown above, one finds here that  $\rho$  has a  broad peak in the second-order region,  in contrast to the first-order region where it undergoes a discontinuous jump at the phase transition. Some remarks are in order:\\

%Fig18

i) At very low temperature, the resistivity increases with decreasing temperature. This behavior can be understood by the freezing of the itinerant spins due to low  $T$:  The  energy of itinerant spins is low, they occupy the low-energy positions in the periodic lattice, it is difficult to move them out by the insufficient thermal energy. They  are somewhat frozen in almost periodic positions; namely a pseudo crystallization occurs.  Note that the increase of resistivity with decreasing $T$ at very low $T$ was observed in many experiments on various materials and is not limited to ferromagnets \cite{Du2007,Wang-Chen,Santos2009,Li}.   This increase of 
$\rho$ with decreasing $T$ in the quantum case has been explained by J. Kondo using a third-order perturbation theory  \cite{Kondo}:  the scattering of $s$-electrons by  $d$-electrons of localized magnetic impurities gives rise to a resistivity minimum at a finite $T$.  We have also found here this minimum of $\rho$  at low $T$  with the classical spin model. The  similarity with the quantum Kondo effect can be explained by the fact that an excited  localized lattice down-spin (in a very small number at low $T$) can be viewed as an impurity which captures nearby conduction up-spins. 

ii) Outside this low-$T$ region, when $T$ increases, the thermal energy progressively unfreezes the itinerant spins. As a consequence, $\rho$ decreases and passes through a minimum (see discussion above). However, at higher $T$, the scattering with the lattice spins is stronger, $\rho$ increases up to the transition temperature.

iii) At the transition temperature, $\rho$ shows a peak. The physical mechanism leading to the peak can be explained: in a previous work \cite{Akabli3}, it was found from our simulations that the peak is due to scattering of the itinerant spins by antiparallel-spin clusters which are numerous in the  transition region. When one gets close to the transition point, the number of clusters of down spins are the most numerous, giving rise to the peak in $\rho$. Note that the "defects" clusters (i. e. clusters of antiparallel spins) have an energy barrier to resist the passage of itinerant spins.  This is also the origin of the extremely long relaxation time in the critical region.

iv) Well above the transition temperature, in the paramagnetic phase, as temperature increases,  clusters of down and up spins will be broken more and more into independent disordered spins, namely spins with zero energy,  itinerant spins move easily on their trajectory, making a decrease of $\rho$ with increasing $T$.

Note that we have also varied the radius $D_1$ to see its effect on $\rho$ at the transition in the present frustrated HCP model. We found the same effect seen in  other antiferromagnets we studied previously \cite{Magnin2,Hoang2011}: at a given temperature, an oscillation of  $\rho$ with varying $D_1$. oscillates slightly with distance. The origin of this oscillation has been analyzed avove in the $J_1-J_2$ model.

\begin{figure}[ht]
\centering
\includegraphics[width=12cm,angle=0]{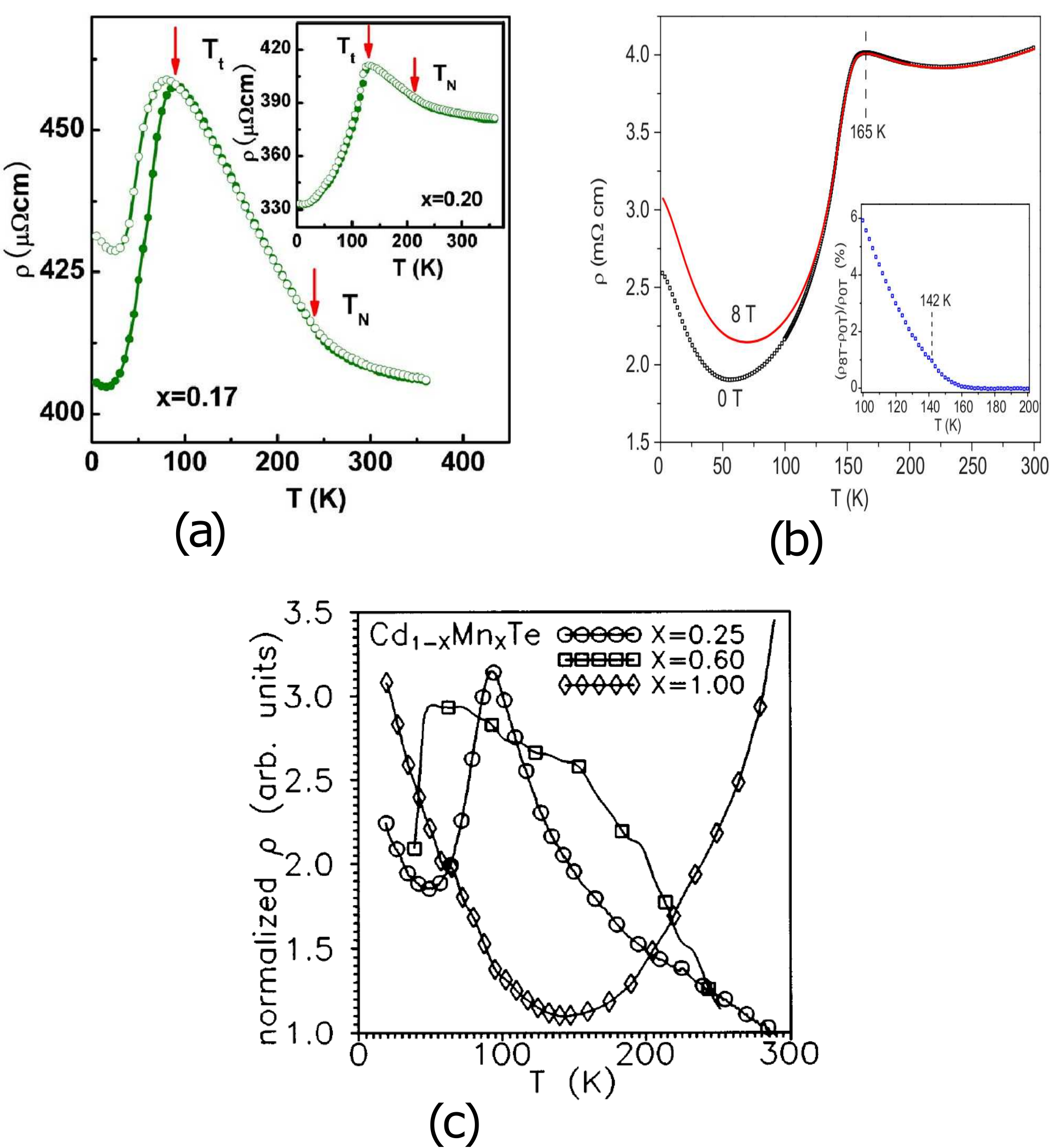}
\caption{Experiments on the resistivity as a function of   $T$ iperformed by (a) by Du et al on $\varepsilon$-(Mn$_{1-x}$Fe$_{x})_{3.25}$Ge antiferromagnets \cite{Du2007}, (b) by McGuire et al. on antiferromagnets LaFeAsO \cite{McGuire2008}, and (c) by Chandra et al. on thin films of Cd$_{1-x}$Mn$_x$Te \cite{Chandra1996}.}
\label{transport_experiment}
\end{figure}

\begin{figure}[ht]
\centering
\includegraphics[width=12cm,angle=0]{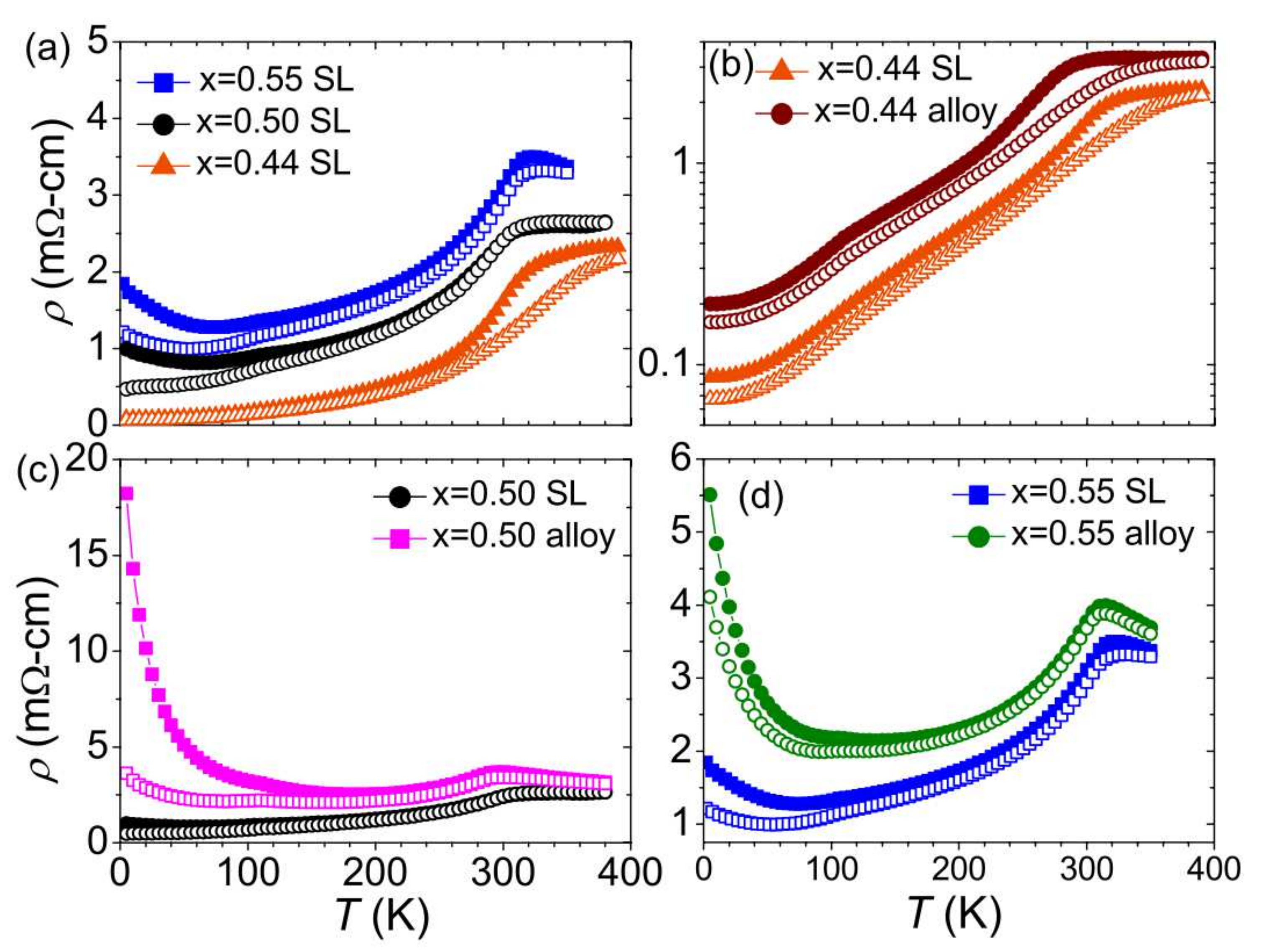}
\caption{Resistivity versus temperature on antiferromagnetic La$_{1-x}$Sr$_x$MnO$_3$. The figures presented are taken from Fig. 7 of \cite{Santos2009}.}
\label{Fig7_Santos}
\end{figure}

Finally, let us look at some experimental data obtained for ferromagnets and antiferromagnets.   Figure \ref{transport_experiment} shows experiments by Du et al. performed on $\varepsilon$-(Mn$_{1-x}$Fe$_{x})_{3.25}$Ge antiferromagnets \cite{Du2007}, experiments by McGuire et al. performed on antiferromagnetic superconductors LaFeAsO \cite{McGuire2008}, by Chandra et al. on thin Cd$_{1-x}$Mn$_x$Te films \cite{Chandra1996}.  Experiments  by Santos et al. on antiferromagnetic La$_{1-x}$Sr$_x$MnO$_3$ \cite{Santos2009} are shown in Fig. \ref{Fig7_Santos}.  We see here that our results on the shape of the spin resistivity are in agreement with these experiments. In the lack of physical data on these experimental materials, we cannot make a quantitative comparison as we did in the MnTe case presented above. 

%Fig19

%Fig20

\section{Conclusion}\label{conclu}
 In this paper, we have reviewed some important works published on the spin resistivity in magnetically ordered systems. We have focused on our works published over the past 15 years using mainly Monte Carlo simulations.  These works were motivated by the absence of Monte Carlo works, even at the present time except ours,  in spite of the fact that this method of simulation has proven to be very efficient when comparing  its results with experimental data. In the case of MnTe where there are sufficient experimental data, we have made a quantitative comparison between experimental and simulated spin resistivities. The agreement between experiments and simulations is excellent. This review therefore aims at promoting this method to study more realistic cases.

As demonstrations, we have used this method to study the spin resistivity in generic ferromagnets and antiferromagnets.  The cases of frustrated systems have also been presented: the $J_1-J_2$ model and the antiferromagnetic HCP lattice.  

Let us summarize the results on the two frustrated systems.

The $J_1-J_2$ model is a simple cubic lattice with Ising spins interacting with each other via NN and NNN antiferromagnetic interactions, $J_1$ and $J_2$ respectively. The GS of this model is determined by the ratio $\eta=J_2/J_1$. We have shown that the GS changes at the critical value $\eta_c=0.25$. For the non-frustrated region in the phase space, namely $\eta<0.25$, the GS is simply composed periodically of two interpenetrated tetrahedra formed by the NNN sites. In the frustrated region, namely  $\eta>0.25$, the GS can be described as composed of one line of spin up, one line of spin down, alternately, in one crystal direction. The degeneracy is three because there is a freedom to choose one direction among three. The total degeneracy is 6 if we count the statesw of reverse spins.  The transition in the frustrated region is theoretically of first order since the present 6-fold GS is equivalent to the $q$-state Potts model with $q=6$. We know that in three  dimensions, the transition of the Potts model is of first order from $q=3$. We have found this directly from the simulation. In the non-frustrated region, namely   $\eta<0.25$,  the transition is found to be of  second order. We have performed MC simulations to obtain $\rho$ of the conduction spins. We found that  $\rho$ displays a broad maximum at the second-order phase transition while it undergoes a discontinuous change at the first-order transition. 

The Ising model on the antiferromagnbetic HCP lattice has been also studied in this review. We assumeed  an in-plane interaction $J_1$ and an inter-plane interaction $J_2$, both antiferromagnetic. We found that  the GS changes at the critical value $\eta_c=0.5$. Below (above) which the spins in the $xy$ planes are ferromagnetic (antiferromagnetic).  The nature of the transition changes  in these two regions:  it is of second order below $\eta_c$ and of first order above  $\eta_c$.  The spin resistivity has been simulated in both regions of $\eta$.  In the second-order region, it shows a broad maximum while in the first-order region, the resistivity $\rho$ makes a discontinuous jump at the transition. This feature is what we also found in  other frustrated spin systems. 

These findings reviewed in this paper show a close relationship between the nature of the phase transition and the shape of the spin resistivity in real materials.  We hope that this review convinces the magnetic community on the use of MC simulations for transport phenomena.


\begin{thebibliography}{999}

\bibitem{Xia} Xia, J.;  Siemons, W.; Koster, G.;  Beasley, M. R.; Kapitulnik, A.  Critical thickness for itinerant ferromagnetism in ultrathin films of SrRuO3, \textit{Phys. Rev. B} \textbf{2009}, \textit{ 79},  R140407.

\bibitem{Lu2009} Lu, C. L.; Chen, X.; Dong, S.;  Wang, K. F.; Cai, H. L.; Liu, J.-M.; Li, D.; Zhang, Z. D. Ru-doping-induced ferromagnetism in charge-ordered La0.4Ca0.6MnO3, \textit{Phys. Rev. B} \textbf{2009}, \textit{79}, 245105.

\bibitem{Du2007} Du, J.;  Li, D.;  Li, Y. B.; Sun, N. K.; Li, J.;   Zhang, Z.D. Abnormal magnetoresistance in
epsilon - [Mn$_{1-x}$Fe$_x$]$_{3.25}$Ge antiferromagnets, Phys. Rev. B {\bf 76}, 094401  (2007).
\bibitem{Zhang} Y. Q. Zhang, Z. D. Zhang and J. Aarts,  Charge-order melting and magnetic phase separation in thin films of Pr0.7Ca0.3MnO3, \textit{Phys. Rev. B} \textbf{2009}, \textit{ 79}, 224422.

\bibitem{Wang-Chen} Wang, X.F.; Wu, T.;  Wu, G.; Chen, H.; Xie, Y. L.; Ying, J. J.;  Yan, Y. J.;  Liu R. H.;  Chen, X.H. Anisotropy in the Electrical Resistivity and Susceptibility of Superconducting BaFe2As2 Single Crystals, \textit{ Phys. Rev. Lett.} \textbf{2009}, \textit{ 102}, 117005.

\bibitem{McGuire2008} McGuire M. \textit{et al.} Phase transitions in LaFeAsO: Structural, magnetic, elastic, and transport properties, heat capacity and M{\"o}ssbauer spectra, \textit{Physical Review B} \textbf{2008}, \textit{ 78}, 094517.

\bibitem{Santos2009} Santos; T. S.;    May, S. J. ;    Robertson, J. L.;   Bhattacharya, A.
Tuning between the metallic antiferromagnetic and ferromagnetic phases of La(1-x)Sr(x)MnO3 near x=0.5 by digital synthesis, \textit{Phys. Rev. B} \textbf{2009},  \textit{ 80}, 155114-155120.
	
	
\bibitem{Matsukura}  Matsukura, F.;  Ohno, H.; Shen, A.; Sugawara, Y. Transport properties and origin of ferromagnetism in (Ga,Mn)As,  \textit{Phys. Rev. B} \textbf{ 1998}, \textit{57}  , R2037.

\bibitem{Stishov} Petrova, A. E.;  Bauer, E. D.; Krasnorussky, V.;  Stishov, S. M.  Behavior of the electrical resistivity of MnSi at the ferromagnetic phase transition,  \textit{Phys. Rev. B} \textbf{2006}, \textit{ 74}, 092401.

\bibitem{Shwerer}  Schwerer, F. C.;   Cuddy, L. J.  Spin-Disorder Scattering in Iron- and Nickel-Base Alloys, \textit{Phys. Rev. B} \textbf{1970}, \textit{2},    1575.

 \bibitem{Kasuya} Kasuya, T. Electrical resistance of ferromagnetic metals, \textit{Prog. Theor. Phys. } \textbf{1956},\textit{16}, 58.

\bibitem{DeGennes} P.-G. de Gennes, P.-G.;  Friedel, J.  Anomalies de r\'esistivit\'e dans certains m\'etaux magn\'etiques, \textit{J. Phys. Chem. Solids} \textbf{1958}, \textit{ 4}, 71.

\bibitem{Fisher}  Fisher, M. E.;  Langer, J. S. Resistive Anomalies at Magnetic Critical Points, \textit{ Phys. Rev. Lett.} \textbf{1968}, \textit{ 20}  , 665.

\bibitem{Kataoka}  Kataoka, M.  Resistivity and magnetoresistance of ferromagnetic metals with localized spins, \textit{Phys. Rev. B} \textbf{2001}, \textit{ 63},  134435.

\bibitem{Zarand} Zarand, G.; Moca, C. P.; Janko, B. Scaling Theory of Magnetoresistance in Disordered Local Moment Ferromagnets, \textit{Phys. Rev. Lett.} \textbf{2005},  \textit{ 94}, 247202.


\bibitem{Fert1}Baibich, M. N.; Broto, J. M.; Fert, A.; Nguyen Van Dau, F.; Petroff, F.; Etienne, P.; Creuzet, G.; Friederich, A.; Chazelas, J. (1988). Giant Magnetoresistance of (001)Fe/(001)Cr Magnetic Superlattices \textit{ Physical Review Letters} \textbf{1988}, \textit{ 61 (21)}, 2472–2475. 

\bibitem{Fert2} Fert, A.  Nobel Lecture: Origin, development, and future of spintronics, \textit{ Rev. Mod. Phys.} \textbf{2008}, \textit{ 80 (4)}, 1517–1530. 

\bibitem{Grunberg} Binasch, G.;  Grünberg, P.;  Saurenbach, F.;  Zinn, W.  Enhanced magnetoresistance in layered magnetic structures with antiferromagnetic interlayer exchange, \textit{Phys. Rev. B} \textbf{1989}, \textit{ 39}, 4828.

\bibitem{Akabli}  Akabli, K.; Diep, H. T.;  Reynal, S. Spin transport in magnetic multilayers, \textit{Journal of Physics: Condensed Matter} \textbf{2007}, \textit{ 19}, 356204.

\bibitem{Akabli2} Akabli, K.; Diep, H. T.  Effects of ferromagnetic ordering and phase transition on the resistivity of spin current, \textit{J. Appl. Phys.} \textbf{2008}, \textit{103}, 07F307.

\bibitem{Akabli3} Akabli, K.;  Diep, H. T. Temperature dependence of the spin resistivity in ferromagnetic thin films: Monte Carlo simulations, \textit{Phys. Rev. B } \textbf{2008}, \textit{ 77}, 165433.

\bibitem{Akabli4} Akabli, K.;  Magnin, Y.;  Oko, M.; Harada, I.;  Diep, H. T. Theory and simulation of spin transport in antiferromagnetic semiconductors: Application to MnTe, \textit{Phys. Rev. B} \textbf{2011}, \textit{ 84}, 024428.

\bibitem{Magnin}  Magnin, Y.;  Akabli, K.;  Diep, H. T.;  Harada, I.  Monte Carlo study of the spin transport in magnetic materials, \textit{Computational Materials Science} \textbf{2010}, \textit{ 49}, S204-S209.

\bibitem{Magnin3} Magnin, Y.;   Hoang, Danh-Tai ;  Diep, H. T. 
Spin Transport in Magnetically Ordered Systems: Effect of the Lattice Relaxation Time,
	\textit{Mod. Phys. Lett. B} \textbf{2011}, \textit{ 25}, 1029.

 \bibitem{Magnin2012} Magnin, Y.;  Diep, H. T.  Monte Carlo Study of Magnetic Resistivity in Semiconducting MnTe , \textit{Phys. Rev. B} \textbf{2012}, \textit{ 85}, 184413.

\bibitem{Magnin2} Magnin, Y.; Akabli, K.;  Diep, H. T.  Spin resistivity in frustrated antiferromagnets,
\textit{Physical Review B} \textbf{2011}, \textit{ 83}, 144406.

\bibitem{Hoang2011}Hoang, Danh-Tai; Magnin, Y.;  Diep, H. T.  Spin Resistivity in the Frustrated $J_1 - J_2$ Model, \textit{Modern Physics Letters B} \textbf{2011}, \textit{ 25}, 937.


\bibitem{DiepFSS} Diep, H. T.  (ed.), {\it Frustrated Spin Systems}, 3rd ed., World Scientific, Singapore (2020).

\bibitem{Haas} Haas, C.  Spin-Disorder Scattering and Magnetoresistance of Magnetic Semiconductors, \textit{Phys. Rev.} \textbf{1968}, \textit{ 168}, 531.

\bibitem{Hohenberg} Hohenberg, P. C.;  Halperin, B. I. Theory of dynamic critical phenomena,  \textit{Rev. Mod. Phys.}  \textbf{1977},  \textit{ 49},  435.

\bibitem{Peczak}  Peczak , P. ;   Landau, D. P.  Monte Carlo study of critical relaxation in the 3D Heisenberg model,  \textit{J. Appl. Phys.}  \textbf{1990},  \textit{ 67}, 5427.

\bibitem{Prudnikov}  Prudnikov, V. V.;  Prudnikov, P. V.;   Krinitsyn, A. S.; Vakilov, A. N.;
Pospelov, E. A.; Rychkov, M. V. Short-time dynamics and critical behavior of the three-dimensional site-diluted Ising model,  \textit{Phys. Rev.  E }  \textbf{2010},  \textit{ 81},  011130, and references therein.

\bibitem{Ngo2007}  Ngo, V. Thanh;   Diep, H. T.  Effects of frustrated surface in Heisenberg thin films, {\it Phys. Rev. B}  \textbf{2007},  \textit{ 75}, 035412.

\bibitem{Pinettes}  Pinettes, C.; Diep, H. T. Phase transition and phase diagram of the J1-J2
Heisenberg model on a simple cubic lattice, {\it J. Appl. Phys.}  \textbf{1998},   \textit{ 83}, 6317.


\bibitem{Hennion} Hennion, B.; Szuszkiewicz, W.; Dynowska, E.;  Janik, E.;  Wojtowicz, T.  Spin-wave measurements on MBE-grown zinc-blende structure MnTe by inelastic neutron scattering,  \textit{Phys. Rev. B}  \textbf{2002},  \textit{ 66},  224426.

\bibitem{Hennion2}  Szuszkiewicz, W.; Dynowska, E.;  Witkowska, B.;  Hennion, B. Spin-wave measurements on hexagonal MnTe of NiAs-type structure by inelastic neutron scattering,
  \textit{ Phys. Rev. B }  \textbf{2006},  \textit{ 73}, 104403.

 \bibitem{Mobasser} Mobasser, S. R.; Hart, T. R. Raman Scattering From Phonons And Magnons In Magnetic Semiconductor MnTe \textit{ Proceed. Society of Photo-Optical Instrumentation Engineers (SPIE),  Conference Series}  \textbf{1985},  \textit{ 524}, 137-144.

\bibitem{Allen} Allen, J. W.;   Locovsky, G.; Mikkelsen Jr., J. C. Optical properties and electronic structure of crossroads material MnTe, \textit{Solid State Commun.} \textbf{1977}, \textit{24}, 367.

\bibitem{Chandra1996} Chandra, S.; Malhotra, L. K.; Dhara, S.;  Rastogi, A. C.  Low-temperature dynamic susceptibility of thin Cd(1-x)Mn(x)Te films,  \textit{Phys. Rev. B}  \textbf{1996},  \textit{ 54},  13694.

\bibitem{Li}  Li, Y. B.;  Zhang, Y. Q.; Sun, N. K.; Zhang, Q.;  Li, D.;  Li, J. ; Zhang, Z. D. Ferromagnetic semiconducting behavior of Mn(1-x)Cr(x)Te compounds,   \textit{Phys. Rev. B}  \textbf{2005},  \textit{ 72}, 193308.

\bibitem{Russe}  Aplesnin, S. S.;  Ryabinkina, L. I.;  Romanova, O. B.; Balaev, D. A.;  Demidenko, O. F.;  Yanushkevich,  K. I.; Miroshnichenko, N. S.  Effect of the orbital ordering on the transport and magnetic properties of MnSe and MnTe,  \textit{Phys. Solid State}  \textbf{2007},  \textit{  49}, 2080-2085. 

\bibitem{Efrem}  Efrem D'Sa, J. B. C.;  Bhobe, P. A.;  Priolkar, K. R.;  Das, A.;  Paranjpe, S. K.; Prabhu, R. P. ; Sarode, P. R. Low-temperature neutron diffraction study of MnTe, 	 \textit{J. Mag. Mag. Mater.}  \textbf{2005},  \textit{ 285 }, 267.

\bibitem{He} He, X.;  Zhang, Y. Q.;  Zhang, Z. D. Magnetic and Electrical Behavior of MnTe(1- x)Sb(x) Alloys,  \textit{ J. Mater. Sci. Technol.}  \textbf{2011},  \textit{ 27}, 64.

\bibitem{samia} Yahyauoi,  S.; Kallel, S.; Diep, H. T.  Magnetic properties of perovskites La(0.7)Sr(0.3)Mn$^{3+}$(0.7)Mn$^{4+}$
(0.3-x) Ti(x)O(3): Monte Carlo simulation versus experiments \textit{Journal of  Magnetism and Magnetic Materials }\textbf{2016}, \textit{416}, 441–448.

\bibitem{Diep1992} Diep, H. T.  First-order transition in the hexagonal-close-packed lattice with vector spins,  \textit{Phys. Rev. B}  \textbf{1992},  \textit{ 45}, 2863.

\bibitem{Hoang2012} Hoang, Danh-Tai;  Diep, H. T. 
Hexagonal-close-packed lattice: Ground state and phase transition,  \textit{Phys. Rev. E}  \textbf{2012},  \textit{ 85}, 041107.



\bibitem{Itakura2003} Itakura, M.  Monte Carlo renormalization group study of the Heisenberg and the XY antiferromagnet on the stacked triangular lattice and the chiral $\phi  4$ model,
 \textit{Journal of the Physical Society of Japan}  \textbf{2003},  \textit{ 72 (1)}, 74-82.

\bibitem{Bekhechi2006} Bekhechi, S.;  Southern, B. W.; Peles,  A.; Mouhanna, D. Short-time dynamics of a family of XY noncollinear magnets,   \textit{Phys. Rev. E}  \textbf{2006},  \textit{74}, 016109.

\bibitem{Ngo2008a}   Ngo, V. Thanh;   Diep, H. T. Stacked triangular XY
antiferromagnets: End of a controversial issue on the phase transition,  \textit{J. Appl. Phys.}  \textbf{2008},  \textit{ 103}, 07C712.

\bibitem{Ngo2008b}   Ngo, V. Thanh;   Diep, H. T.  Phase transition in Heisenberg stacked triangular antiferromagnets: End of a controversy,  \textit{Phys. Rev. E}  \textbf{2008},  \textit{ 78}, 031119.
% SAFT
\bibitem{Delamotte2005}   Delamotte, B. ;Mouhanna, D.;  Tissier, M.  chapter {\it Renormalization Group Approaches to Frustrated Magnets in D=3}  in Ref. \cite{DiepFSS}.

\bibitem{Diep-Kawamura} Diep, H. T.;  Kawamura, H. First-order phase transition in the fcc Heisenberg antiferromagnet, \textit{Phys. Rev. B} \textbf{1989}, \textit{ 40}, 7019.

\bibitem{Blankschtein1984} Blankschtein, D.;  Ma, M;  Berker, A. N. Fully and partially frustrated simple-cubic Ising models: Landau-Ginzburg-Wilson theory,  \textit{Phys. Rev. B} \textbf{1984}, \textit{ 30}, 1362.


\bibitem{Ngo2010} Ngo, V. Thanh;   Hoang, D. Tien;   Diep, H. T.  First-order transition in the XY model on a fully frustrated simple cubic lattice,  \textit{Phys. Rev. E}  \textbf{2010},  \textit{ 82} ,  041123.


\bibitem{Ngo2011a} Ngo, V. Thanh;   Hoang, D. Tien;   Diep, H. T.   Phase Transition In The Heisenberg Fully-Frustrated Simple Cubic Lattice, \textit{Mod. Phys. Lett. B} \textbf{2011}, \textit{  25}, 929.

\bibitem{Ngo2011b}  Ngo, V. Thanh;   Hoang, D. Tien;   Diep, H. T.  Flat energy-histogram simulation of the phase transition in an Ising fully frustrated lattice,  \textit{J. Phys.: Cond. Matt.} \textbf{2011}, \textit{ 23}, 226002.


 \bibitem{Diep1989} Diep, H. T.  Magnetic transitions in helimagnets, \textit{Phys. Rev. B} \textbf{1989}, \textit{ 39}, 397.

\bibitem{Kondo} See "Kondo Effect - 40 Years after the Discovery" - Special Issue of the Journal of the Physical Society of Japan \textbf{2015},  \textit{74}, Issue 1, with reviews from world top scientists: Jun Kondo, Philippe Nozi\`eres, A. C. Hewson, Yukihiro Shimizu and Osamu Sakai, Ian Affleck amongst others.





%%%%%%%%%%%%%%%




\end{thebibliography}
\end{document}